\documentclass{LMCS}

\def\doi{8 (1:27) 2012}
\lmcsheading%
{\doi}
{1--35}
{}
{}
{Mar.~\phantom02, 2010}
{Mar.~26, 2012}
{}

\usepackage{ifthen}
\usepackage{amssymb} 
\usepackage{amsmath} 
\usepackage{amscd}
\usepackage{xspace} 
\usepackage{graphicx}
\usepackage{enumerate,hyperref}

\newcommand{\thmcite}[2]{{\normalfont\textit{#1}\,#2}}
\newcommand{\tw}{\operatorname{tw}}
\newcommand{\ar}{\textit{ar}}

\newcommand{\vrt}[1]{V(#1)}

\newcommand{\vrtB}{\vrt{B}}
\newcommand{\vrtG}{\vrt{G}}

\newcommand{\vrtP}{\vrt{P}}
\newcommand{\vrtT}{\vrt{T}}

\newcommand{\edge}[1]{E(#1)}

\newcommand{\edgeG}{\edge{G}}

\newcommand{\tplX}{\tpl{X}}

\newcommand{\isom}{\cong}

\newenvironment{Notation}{\par\medskip\noindent\textbf{Notation.\ }}{\par\medskip}

\newenvironment{cenv}{\begin{list}{}{%
      \setlength{\labelwidth}{1.5em}%
      \setlength{\leftmargin}{\labelwidth}%
      \addtolength{\leftmargin}{\labelsep}%
      \setlength{\listparindent}{0em}%
      \setlength{\topsep}{10pt}%
      \setlength{\itemsep}{5pt}%
      \setlength{\parsep}{0pt}%
    }
  }{
  \end{list}
}

\newcounter{claimcounter}

\newcommand{\st}{\mathrel :}

\let\oldtheta\theta
\renewcommand{\theta}{\vartheta}

 \newcommand{\BBB}{{\mathcal B}}
\newcommand{\CCC}{{\mathcal C}} 
 
\newcommand{\GGG}{{\mathcal G}} 
\newcommand{\III}{{\mathcal I}} 
 \newcommand{\LLL}{{\mathcal L}}
 
 \newcommand{\PPP}{{\mathcal P}}
\newcommand{\QQQ}{{\mathcal Q}}

\newcommand{\Vp}{\mathcal{V}}
\newcommand{\Hp}{\mathcal{H}}

\renewcommand{\AA}{{\mathfrak A}} \newcommand{\BB}{{\mathfrak B}}
 \newcommand{\DD}{{\mathfrak D}}
 
\newcommand{\GG}{{\mathfrak G}}

\newcommand{\N}{{\mathbb N}} 
 \newcommand{\Z}{{\mathbb Z}}

\renewcommand{\emptyset}{\varnothing} \renewcommand{\phi}{\varphi}

\newcommand{\Oof}{\mathcal{O}}

 \newcommand{\PTIME}{\mbox{\sc
Ptime}} 
\newcommand{\NP}{\mbox{\rm NP}} 
\newcommand{\PSPACE}{\mbox{\sc Pspace}\xspace}

\newcommand{\EXPTIME}{\mbox{\sc Exptime}}
\newcommand{\NEXPTIME}{\mbox{\sc NExptime}}

\newcommand{\Pow}{\ensuremath \PPP}

 \renewcommand{\theta}{\vartheta}

\newcommand{\FO}{\textup{{FO}}\xspace}
\newcommand{\MSO}{\textup{{MSO}}\xspace}

\newcommand{\tpl}[1]{\overline{#1}} 
\newcommand{\tpla}{\tpl{a}}
\newcommand{\tplb}{\tpl{b}}

 \newcommand{\tplx}{\tpl{x}}

\newcommand{\tplY}{\tpl{Y}}

\newcommand{\look}{\begin{proof}} 
\newcommand{\hx}{\end{proof}}

\newcommand{\Iff}{\Longleftrightarrow} 
 
 \renewcommand{\phi}{\varphi}
\renewcommand{\theta}{\vartheta} \renewcommand{\epsilon}{\varepsilon}

\newcommand{\minor}{\preceq}

\newcommand{\ExQed}{\hspace*{1pt}\nobreak\hfill$\dashv$}



\renewenvironment{Proof}{\proof}{\qed}

\newcommand{\MC}{\textup{MC}}
\newcommand{\pMC}{\textup{$p$-MC}}

\newcommand{\sigmai}{\sigma_{\textit{inc}}}
\newcommand{\sigmaw}{\sigma_{\textit{G}}}
\newcommand{\sigmacol}{\sigma_{\textit{col}}}
\newcommand{\sigmaord}{\sigma_{\textit{ord}}}

\newcommand{\phiuniv}{\phi_{\textit{univ}}}
\newcommand{\phivalid}{\phi_{\textit{valid}}}

\newcommand{\sodp}{\textit{set-o-dis-path}}
\newcommand{\maxpath}{\textit{maxpath}}
\newcommand{\phiendpoint}{\textit{ep}}
\newcommand{\uniqueedge}{\textit{uni-edge}}
\newcommand{\pAth}{\textit{path}}

\newcommand{\wall}[1]{}

\begin{document}

\title{On the Parameterized Intractability of Monadic Second-Order
   Logic}

\author[S.~Kreutzer]{Stephan Kreutzer}

\address{School for Electrical Engineering and Computer Science, Technical University Berlin, Sekr. TEL 7-1, Ernst-Reuter-Platz 7, 10587 Berlin, Germany} 
\email{stephan.kreutzer@tu-berlin.de} 
\thanks{Research supported by DFG
    grant KR 2898/1-3. Part of this work was done
    while the author participated at the workshop "Graph Minors"
    at Banff International Research Station, October 2008.}


\maketitle

\begin{abstract}
  \noindent{}One of Courcelle's celebrated results states that if $\CCC$ is a
  class of graphs of bounded tree-width, then model-checking for
  monadic second order logic $(\MSO_2)$ is fixed-parameter tractable
  (fpt) on $\CCC$ by 
  linear time parameterized algorithms, where the parameter is the
  tree-width plus the size of the formula. 
  An immediate question is
  whether this is best possible or whether the result can be
  extended to classes of unbounded tree-width. 

  In this paper we show that in terms of tree-width, the theorem
  cannot be extended much further.  More specifically, we show that if
  $\CCC$ 
  is a class of graphs which is closed under colourings and satisfies
  certain constructibility conditions and is such that the tree-width of
  $\CCC$ is not bounded by $\log^{84} n$ then $\MSO_2$-model
  checking is not fpt unless
  \textsc{Sat} can be solved in sub-exponential time. If the
  tree-width of $\CCC$ is not poly-logarithmically bounded, then $\MSO_2$-model
  checking is not fpt unless all problems in
  the polynomial-time hierarchy can be solved in sub-exponential time.  
\end{abstract}

\section{Introduction}

\noindent Classical logics such as first-order or fragments of second-order
logic have played a crucial role in the development and analysis of query or
specification languages in database theory, formal language theory
and many other areas. In these application areas, computational logic problems
such as satisfiability and model checking occur frequently and much
effort has gone into analysing the complexity of these computational
tasks. 

In this paper we are mostly concerned with model checking for monadic
second-order logic $(\MSO_2)$, the extension of first-order logic by
quantification over sets of elements (i.e.~vertices and edges). The model-checking
problem for $\MSO_2$ is the 
problem to decide for a given structure and a formula whether the
formula is true in the structure. A reduction from the
\PSPACE-complete \emph{quantified boolean formula}-problem (QBF) 
immediately shows that 
the model-checking problem for first-order and monadic second-order logic
is \PSPACE-hard. In fact the problems are \PSPACE-complete \cite{Vardi82}.
The problem even remains \PSPACE-complete on a fixed structure with
only two elements, showing that the high complexity is already
generated by the formula alone.

However, especially in a database context, where a formula specifies
a query and the structure is the database, it can usually be assumed
that the formula is reasonably small whereas the database is very
large. Vardi \cite{Vardi82} therefore proposed the concept of
\emph{data complexity} which is the complexity of model-checking
against a fixed formula. For first-order logic, it can be shown that
the data complexity is always polynomial time whereas for monadic
second-order logic the model checking problem can already be \NP-hard
for a fixed formula. See e.g.~Section~\ref{sec:mso} for an example
defining the \NP-complete 3-colourability problem. 
However, even for first-order logic, where the model-checking problem
has polynomial time data complexity, the algorithms witnessing this
usually run in time $|\AA|^{\Oof(|\phi|)}$ and hence in time exponential in
the formula. As the database $\AA$ was assumed to be huge, this is
unacceptable even for relatively small formulas $\phi$. 

A more refined analysis of the model-checking complexity separating
the complexity with respect to the formula from the complexity in
terms of the database is offered by the framework of
\emph{parameterized complexity} \cite{DowneyFel98,FlumGro06}.
In this framework, the input to a model-checking problem again
consists of a pair $(\AA, \phi)$, where $\AA$ is a finite structure
and $\phi$ is a formula, but now we declare $|\phi|$ as the
\emph{parameter}. We call the problem \emph{fixed parameter tractable}
(fpt), if it can be solved in time $f(|\phi|)\cdot |\AA|^c$, where $f$
is a computable function and $c$ a constant. Hence, we allow arbitrary
amount of time with respect to the size of the formula but only fixed
polynomial time in the size of the structure. The problem is in the
parameterized complexity class XP if it can be solved in time
$|\AA|^{f(|\phi|)}$. 
The class FPT of all
fixed-parameter tractable problems is the parameterized analogoue of
polynomial time in classical complexity as model of tractable
computation. The class XP takes over the role of  exponential time in
classical complexity. 

Model-checking problems have received particular
attention in the context of parameterized complexity. See
e.g.~\cite{PapadimitriouY99} for a discussion on query complexity in
databases theory with respect to the framework of parameterized complexity.

As the example of an \MSO-formula defining
$3$-colourability shows, on general graphs model-checking for monadic
second-order logic is not fixed-parameter tractable unless $\PTIME=\NP$.
However, fixed-parameter tractability can be retained by restricting
the class of admissible structures, for instance to words or
trees. Studying properties and complexity results for monadic
second-order logic on restricted classes of structures has a very long
tradition in computer science, going back to by now classical results
by B\"uchi, Rabin, Doner, Thatcher and  Wright that on words and trees any
formula of monadic second-order logic is equivalent to a word- or
tree-automaton and hence, in terms of parameterized complexity, the
model checking problem on such structures becomes fixed-parameter
tractable as follows: given a tree $T$ and a monadic second-order
logic formula $\phi$, we first convert $\phi$ into an equivalent
tree-automaton, which is costly but only depends on $|\phi|$, and then
let the automaton run on the tree $T$ to verify $T\models \phi$. The
latter runs in linear time in the size of $T$, hence the whole
model checking algorithm runs in time $f(|\phi|)\cdot |T|$ and is
therefore fixed-parameter linear. 

The observation that even such a powerful logic as monadic
second-order logic becomes fixed-parameter linear on trees has been
used in numerous contexts and applications. In database theory in particular, it has
influenced the development of query languages for XML
databases, a database model designed for data integration on the
web. XML databases are tree-like, in the sense that their skeleton is
a tree (but there may be additional \emph{references} creating edges
violating the tree-property). The tree-structure and unbounded depth of XML databases
necessitates new query languages such as XPath and others which allow
to navigate in the tree, especially along paths from a node to its
direct or indirect successors. In this context, monadic second-order
logic has played the role of a yardstick as \MSO-queries can be
evaluated in linear time yet prove to be very expressive.

To be able to fully explore the potential of logics such as monadic
second-order logic or first-order logic for future applications in
databases and elsewhere, a thorough understanding of the structural
properties of models that allow for tractable model-checking would
prove most useful. 

Ideally, for common logics $\LLL$ such as $\FO$
or variants of $\MSO$, we aim at identifying a property $P$
such that the 
parameterized model-checking problem for $\LLL$ becomes tractable on a
class $\CCC$ of databases (or logical structures) if, and only if,
$\CCC$ has the structural property $P$ (under reasonable complexity
theoretical assumptions).

There may not always exist such a property that
precisely captures tractability for a logic, and sometimes we may have
to compromise and impose further restrictions on the class $\CCC$,
such as closure under sub-structures. But any reasonably precise
characterisation 
would have great potential for future use of these logics in query and
specification languages. 

In this paper we establish a first characterisation in this sense of
monadic second-order logic ($\MSO_2$), or
more generally \emph{guarded second-order logic}.

In 1990, Courcelle proved a fundamental result stating that 
every property of graphs definable in monadic second-order logic
($\MSO_2$) can be decided in 
linear time on any class $\CCC$ of structures of bounded
tree-width (see below for a definition of tree-width). 
Besides the applications to logic outlined above,
Courcelle's theorem has had significant impact on the theory of
parameterized problems on graphs.
In the design of efficient algorithms
on graphs, it can often be used as a simple way of
establishing that a property can be solved in linear time on graph
classes of bounded tree-width. 
Furthermore, results
such as Courcelle's theorem, usually called \emph{algorithmic meta-theorems}, lead
to a better understanding how far certain algorithmic techniques range
and establish general upper bounds for the parameterized complexity
of a wide range of problems.  
See \cite{Grohe07,Kreutzer11,GroheK11} for recent surveys on algorithmic
meta-theorems.

From a logical perspective, Courcelle's theorem establishes a
sufficient condition for tractability of $\MSO_2$ formula
evaluation on classes $\CCC$ of structures: whatever the class
$\CCC$ may
look like, if it has bounded tree-width, then $\MSO_2$-model checking
is tractable on $\CCC$. 
An obvious question is how tight
Courcelle's theorem is, 
i.e.~whether it can be extended to classes of unbounded tree-width and
if so, how ``unbounded'' the tree-width of graphs in the class can be in
general. 
This question is the main motivation for the work reported here. 

In this paper we establish an intractability result
by showing that in its full generality, Courcelle's theorem can not be
extended much further to classes of unbounded tree-width. 
Throughout the paper we consider structures over a binary signature
$\sigma := \{ R_1, \dots, R_k, P_1, \dots, P_l, c_1, \dots, c_k\}$,
where the $R_i$ are binary relation symbols, the $P_i$ are unary and
the $c_i$ are constants. We require that $\sigma$ contains at least
two binary and two unary relation symbols. See Section~\ref{sec:mso}
for details. To give an example, an XML database over a fixed schema
can naturally be modelled by a structure over a binary signature where
each axis label yields a binary relation in the obvious way. Another
intuitive way of looking at binary structures is to view them as
coloured graphs, where the binary relations correspond to edge colours
and the unary relations to vertex colours. As it helps simplifying the
presentation, we will adapt this way of looking at binary structures.

To state our main result, we first need some notation.
\begin{defi}\label{def:gaifman}
  Fix a binary signature $\sigma$ as before. The Gaifman-graph
  $\GGG(\AA)$ of a $\sigma$-structure $\AA$ is the graph with the same
  universe as $\AA$ and an edge $\{ a, b\}$ if there is a binary
  $R_i\in \sigma$ such that $(a,b)\in R_i^{\AA}$ or $(b, a)\in
  R_i^{\AA}$. A class $\CCC$ of $\sigma$-structures is said to be
  closed under colourings, if whenever $\AA\in\CCC$ and $\GGG(\AA)
  \isom \GGG(\BB)$ then $\BB\in\CCC$.

\end{defi}Informally, whenever two $\sigma$-structures only differ
in the colours of edges and vertices, then they both belong to $\CCC$
or both do not.

\begin{defi}\label{def:strongly}
  Let $\sigma$ be a binary signature. Let $f\st \N\rightarrow \N$ be a
  function and $p(n)$ be a polynomial.

  The tree-width of a class $\CCC$ of $\sigma$-structures is
  \emph{$(f, p)$-unbounded}, if for all $n\geq 0$ 
  \begin{enumerate}[(1)]
  \item there is a graph $G_n\in\CCC$ of tree-width $\tw(G_n)$ between $n$ and
    $p(n)$ such that $\tw(G_n) > f(|G|)$ and 
  \item  given
    $n$, $G_n$ can be constructed in time $2^{n^\epsilon}$, for
    some $\epsilon<1$.
  \end{enumerate}
  The
  tree-width of $\CCC$ is \emph{poly-logarithmically unbounded} if there
  are polynomials $p_i(n)$, $i\geq 0$, so that $\CCC$ is $(\log^i,
  p_i)$-unbounded for all $i$.
\end{defi}

\noindent See Section~\ref{sec:prelims} for a definition of tree-width and
related concepts.
Essentially, the first condition ensures that there are
not too big gaps between the tree-width of graphs witnessing that the
tree-width of $\CCC$ is not bounded by $f(n)$. The second condition
ensures that we can
compute such witnesses efficiently, i.e.~in time polynomial in their
size. We will see below why these conditions are needed. The following
is the main result of the paper.

\begin{thm}\label{theo:main}
  Let $\sigma$ be a binary signature with at least two binary and two
  unary relation symbols. 
  Let $\CCC$ be a class of
  $\sigma$-structures closed under colourings. 
  \begin{enumerate}[\em(1)]
  \item If the tree-width of $\CCC$ is 
    poly-logarithmically unbounded then $\MC(\MSO_2,\CCC)$ is not
    in XP and hence not fixed-parameter tractable unless all problems in $\NP$ (in fact,
    all problems in the polynomial-time hierarchy) can be
    solved in sub-exponential time.
  \item If the tree-width of $\CCC$ is $(\log^c, p)$-unbounded, for
    some  $c>d\cdot 84$ and polynomial $p$ of degree $d$, 
    then $\MC(\MSO_2, \CCC)$ is not in XP and hence not fixed-parameter tractable unless
    \textsc{Sat} can be solved in sub-exponential time. 
  \end{enumerate}
\end{thm}

See Section~\ref{sec:mso} for a precise definition of $\MSO_2$ over
structures and Section~\ref{sec:complexity} for a definition of FPT
and XP. 

Essentially, as far as classes closed under colourings are concerned,
if the tree-width of a class of graphs is not logarithmically bounded,
then it has intractable $\MSO_2$ model-checking.
In this sense the theorem shows that tractability results as general
as Courcelle's are not possible for classes of more than
logarithmic tree-width. The restriction to classes closed under
colourings is obviously a real restriction and it is possible that
there are very special classes of $\sigma$-structures of tree-width not bounded by
$\log^{84} n$ but with tractable model-checking. However, the
usefulness of monadic second-order logic and tractability results such
as Courcelle's theorem lie in their general applicability as
specification and query languages. After all, we want a
query language to be tractable on all databases of a certain
structure, and not just if they have the right labels on their
axes. And our result shows that beyond logarithmic tree-width, $\MSO$
no longer fulfills this promise. 

 Compared to
Courcelle's theorem, there is a
gap between constant tree-width to which Courcelle's theorem applies
and tree-width not bounded by $\log^{84} n$ to which our theorem applies. 
The bound $c>d\cdot 84$ can be improved to $c>d\cdot 48$, see
Section~\ref{sec:conclusion}, and conceivably can be improved further.  
However, Makowsky and Mari\~no \cite{MakowskyM03} exhibit a class of
graphs whose tree-width is only bounded by $\log n$, i.e.~it is
$(\log^c n, n)$-unbounded for all $c<1$, but where $\MSO_2$
model-checking is tractable. It is easily seen that the closure of
this class under colourings still admits tractable $\MSO_2$
model-checking. Hence, there is no hope to extend our result to
classes of tree-width less than logarithmic. 

Let us give some applications of the theorem. For $c>0$ let $\CCC_c$ be the
class of all graphs $G$ of tree-width at most $\log^c |G|$. Then the 
closure under colourings has intractable $\MSO_2$ model-checking, if
$c>84$. Similarly, intractability follows for the class of planar 
graphs of tree-width at most $\log^c n$, as colours
in this class can easily be encoded.
All these examples show that Courcelle's theorem can not be extended
to classes of graphs with only poly-logarithmic or
a $\log^{c} n$ bound on the tree-width, for $c> 84$. 

Following Courcelle's theorem, a series of algorithmic meta-theorems
for first-order logic on planar graphs \cite{FrickGro01}, (locally)
$H$-minor-free graphs \cite{FlumGro01,DawarGroKre07} and various other
classes have been obtained. Again, no deep lower bounds,
i.e.~intractability conditions, are known (see~\cite{Kreutzer11} for
some bounds and 
\cite{Grohe07,Kreutzer11,GroheK11} for recent surveys of the topic). The aim of
this paper is to initiate a thorough study of sufficient conditions
for intractability in terms of structural properties of input
instances.

\noindent\textbf{Related work. }
Lower bounds for the complexity of monadic second-order logic for
specific classes of graphs have been considered in the literature
before. In \cite{MakowskyM03}, Makowsky and Mari\~no show that if a
class of graphs has unbounded tree-width and is closed under
topological minors then model-checking for $\MSO_2$ is not
fixed-parameter tractable unless $P = \NP$. 

In \cite{CourcelleMakRot00}, Courcelle et al. show that unless
$\EXPTIME = \NEXPTIME$, model-checking for $\MSO_2$ is not
fixed-parameter tractable on the class of complete graphs. 

More closely related to the result reported here, Grohe
\cite[Conjecture 8.3]{Grohe07} conjectures that
$\MSO$-model checking is not fixed-parameter tractable on 
any class $\CCC$ of graphs which is closed under taking subgraphs and
whose tree-width is not poly-logarithmically bounded, i.e.~there are no
constants $c,d$ such that $\tw(G) \leq d\cdot\log^c |G|$ for all $G\in \CCC$.

Grohe's conjecture was affirmed  in
\cite{KreutzerTaz10,KreutzerTaz10b} with respect to certain technical conditions
similar to the notion of $(f, p)$-unboundedness defined above. It was proved that if $\CCC$ is
closed under sub-graphs and its tree-width is $(\log^c, p)$-unbounded
for some small constant $c$, then $\MSO_2$ model-checking is not fpt
on $\CCC$ unless SAT can be solved in sub-exponential time. The proof
of this result is considerably more complex and much more technical than the proof reported
here, especially in its combinatorial core.

It is worth noting that the two results are somewhat incomparable. In
particular, closure under sub-structures in this
context is a stronger requirement than it might seem at first
sight: while tree-width is preserved by taking sub-graphs,
logarithmic or poly-logarithmic tree-width is not. I.e., a sub-graph
of a graph of tree-width at most $k$ also has tree-width at most $k$,
but if $G$ has tree-width at most logarithmic in its order, this may
not be the case for sub-graphs. Hence, the results in
\cite{KreutzerTaz10b} are much more
restrictive in this sense than our result here. On the other hand, they
do not require closure under colourings and are therefore
much more general in this aspect.

\smallskip

\noindent\textbf{Organisation. } We fix our notation and review
the graph theoretical notions we need in Section~\ref{sec:prelims}. 
Monadic second-order logic is defined in Section~\ref{sec:mso} and its
complexity is reviewed in Section~\ref{sec:complexity}.
We give an informal and intuitive presentation of the main proof idea
in Section~\ref{sec:overview}. The proof is presented in full detail
in Sections~\ref{sec:pseudo-walls} to~\ref{sec:main}. We conclude in Section~\ref{sec:conclusion}.
\smallskip

\noindent\textbf{Acknowledgements. } I would like to thank Mark Weyer for
pointing out that the result proved here readily extends to problems
in the polynomial time hierarchy. Many thanks also to the referees for
many helpful comments improving the presentation of the paper.

\section{Preliminaries}
\label{sec:prelims}

In this section we fix our notation and review concepts from graph
theory needed below.

\subsection{General Notation. }
If $M$ is a set we write $\Pow(M)$ for the set of all subsets of
$M$. If $M, N$ are two sets, we define $M\dot\cup N$ as the
\emph{disjoint union} of $M$ and $N$, obtained by taking the union of
$M$ and a copy $N'$ of $N$ disjoint from $M$. We also apply this
notation to graphs and other structures for which a union operation is
defined.

We write $\Z$ for the set of integers and $\N$ for the set of
non-negative integers. 

\subsection{Graphs and Colourings}

We will use standard notation from graph theory and refer to 
\cite{Diestel05} for background on graphs and details on the graph
theoretical concepts introduced in this section.

All graphs in this paper are finite, 
undirected and simple, i.e.~without multiple edges or loops. 
We write $V(G)$ for the set of vertices and $E(G)$ for the
set of edges in a graph $G$. We will always assume that
$\vrtG\cap\edgeG =\emptyset$.

The \emph{order} $|G|$ of a graph is defined as $|V(G)|$ and its
\emph{size} $||G||$ as the number of edges. 

For $l\geq 1$ we denote the \emph{$l$-clique}, the  complete graph on
$l$ vertices,  by $K_l$. 
 
A graph $H$ is a \emph{sub-division} of $G$ (a \emph{$1$-subdivision}) if $H$ is obtained from $G$ by
replacing edges in $G$ by paths of arbitrary length (of length $2$,
resp.). $H$ is a
\emph{topological minor} of $G$ if a subgraph $G'\subseteq G$ is
isomorphic to a sub-division of $H$. 

A graph $H$ is a \emph{minor} of
$G$ if it can be obtained from a sub-graph $G'\subseteq G$ by
contracting edges. An equivalent, sometimes more intuitive,
characterisation of the minor relation can be obtained using the
concept of \emph{images}. $H$ is a minor of $G$ if there is a map
$\mu$ mapping each $v\in
V(H)$ to a tree $\mu(v) \subseteq G$ and each edge $e\in E(H)$ to an
edge $\mu(e)\in E(G)$ such that if $u\not= v\in
V(H)$ then $\mu(v) \cap \mu(u) = \emptyset$ and if $\{u,v\}\in E(H)$ then
$\mu(\{u,v\}) = \{ u', v'\}$ for some $u' \in V(T_u)$ and
$v'\in V(T_v)$. $\mu$ is called the \emph{image map} and
$\bigcup_{v\in V(H)} \mu(v) \cup \bigcup_{e\in E(H)} \mu(e) \subseteq
G$ is called the \emph{image} of $H$ in $G$.
It is not difficult to see that $H\minor G$ if, and
only if, there is an image of $H$ in $G$.

\wall{The following lemma will be useful later. 
\begin{lem}\label{lem:top-minor}
  If $H$ has maximum degree $\leq 3$ and $H\minor G$ then $H$ is a
  topological minor of $G$. 
\end{lem}

To see this, let $H$ be of degree $\leq 3$ and let $\mu$ be an image
map of $H$ in $G$. Obviously, for $v\in V(H)$, if  the tree $\mu(v)$
contains a leaf which is not the endpoint of an
edge $\mu(e)$, then we can remove this leaf from $\mu(v)$ and still
have an image of $H$ in $G$. Hence, we can assume that all $\mu(v)$
have at most $3$ leaves and therefore only on vertex of degree
$>2$. The image of $H$ in $G$ witnessed by $\mu$ is therefore
isomorphic to a sub-division of $H$. }

Let $G$ be a graph and $A,B \subseteq V(G)$. A set $S\subseteq V(G)$
is an \emph{A-B-separator} if there is no path in $G\setminus S$ from
a vertex in $A$ to a vertex in $B$. An \emph{A-B-path} $P\subseteq G$
is a path in $G$ with one endpoint in $A$ and the other in $B$.

\begin{thm}[Menger]\label{thm:menger}
  Let $G$ be a graph and $A, B \subseteq V(G)$. The minimal
  cardinality $|S|$ of an A-B-separator $S \subseteq V(G)$ is equal to
  he maximum number of vertex disjoint A-B-paths in $G$.
\end{thm}

Finally, we will be using the concept of intersection graphs.

\begin{defi}\label{def:intersection-graph}
  Let $G$ be a graph and $\PPP, \QQQ$ be two sets of pairwise disjoint
  paths in $G$. The \emph{intersection
  graph} $\III(\PPP, \QQQ)$ is defined as the graph with vertex set
$\PPP \dot\cup \QQQ$ where $P, Q$ are adjacent if, and only if, $P\cap
Q \not=\emptyset$.
\end{defi}

\subsection{Tree-Width and Obstructions}

Tree-width is a measure of similarity of graphs to being a tree that was
introduced by Robertson and Seymour in their graph minor project
(\cite{GM-series}), even though equivalent concepts have been studied
under different names before \cite{Halin76,Rose70}. 

\begin{defi}
  A \emph{tree-decomposition} of a graph $G$ is a pair $(T,
  (B_t)_{t\in\vrtT})$ where $T$ is a tree and $B_t\subseteq \vrtG$
  such that
  \begin{enumerate}[(1)]
  \item for all $v\in \vrtG$, the set $\{ t\in\vrtT \st v\in B_t
    \}$ is non-empty and connected in $T$ and
  \item for every edge $e := \{ u,v\} \in\edgeG$ there is a
    $t\in\vrtT$ such that $u,v\in B_t$.
  \end{enumerate}
  The \emph{width} of a tree-decomposition is $\max_{t\in\vrtT}
  |B_t|-1$ and the \emph{tree-width} $\tw(G)$ of a graph $G$ is the
  minimal width of any of its tree-decompositions.

  A class $\CCC$ of graphs has \emph{bounded tree-width} if there is a
  constant $c\in \N$ such that $\tw(G) \leq c$ for all $G\in\CCC$.
\end{defi}

Many natural classes of graphs are found to have bounded tree-width,
for instance series-parallel graphs or control-flow graphs of goto-free
C programs \cite{Thorup98}, and many 
generally NP-hard problems can be solved efficiently on graph
classes of small tree-width. This is witnessed in particular by
Courcelle's Theorem~\ref{theo:courcelle} below.

In this paper, we will mostly be concerned with
graphs of large tree-width and structural information we can gain
about a graph once we know that its tree-width is large. This leads to
the concept of \emph{obstructions}, i.e. structures we can find in any
graph of large enough tree-width. In this paper, we will use two such
obstructions, \emph{brambles} and \emph{grids\wall{/walls}}.

\begin{defi}
  Let $G$ be a graph. Two subgraphs $X,Y\subseteq G$ \emph{touch} if
  $X\cap Y\not=\emptyset$ or there is an edge in $G$ \emph{linking} $X$ and
  $Y$, i.e.~with one endpoint in $X$ and the other in $Y$. A \emph{bramble} in $G$ is a set $\BBB$ of pairwise
  touching connected subgraphs of $G$.  
  A set $S\subseteq V(G)$ is a \emph{cover} for
  $\BBB$ if $S\cap V(B)\not=\emptyset$ for all $B\in\BBB$. The
  \emph{order} of $\BBB$ is the minimum cardinality of a cover of
  $\BBB$. The \emph{size} of  $\BBB$ is the number $|\BBB|$ of sets in $\BBB$. 
\end{defi}

Brambles provide a dual characterisation of tree-width as shown in the
following theorem. 

\begin{thm}[\cite{SeymourTho93}]\label{theo:bramble} 
  A graph $G$ has treewidth at
  least $k$ if, and only if, $G$ contains a bramble of order at least $k+1$.
\end{thm}

\begin{figure}[t]
  \includegraphics[height=2cm]{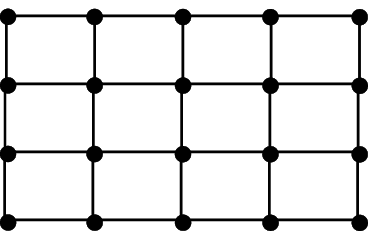} 
  \centering
\wall{  \caption{a) A $(4\times 5)$-grid and b) a $(5\times
    6)$-wall.}}
\caption{A $(4\times 5)$-grid.}
  \label{fig:wall}
\end{figure}

Brambles will form the basis of our algorithmic part below. However,
often it is much easier to work with another obstruction, known as \emph{grids}.
A \emph{$(k\times l)$-grid} $G_{k\times l}$ is a graph as in
Figure~\ref{fig:wall}. Formally, $G_{k\times l}$ is defined as the
graph with 
\begin{eqnarray*}
  V(G_{k\times l}) &:=& \{ (i,j) \st 1\leq i\leq k, 1\leq j\leq
  l \}\text{ and }\\
  E(G_{k\times l}) &:=& \{ \{ (i,j), (i', j') \} \st |i-i'| + |j-j'| = 1
  \}.
\end{eqnarray*}

Grids play a very special role in connection with tree-width as every
graph of large tree-width contains a large grid as minor.

\begin{thm}[Excluded Grid Theorem \cite{GMV,RobertsonSeyTho94}] \label{thm:excl-grid}
  There is a function  $f \st \N\rightarrow \N$ such that any graph
  of tree-width at least $f(k)$ contains a $(k\times k)$-grid as a minor.
\end{thm}

Unfortunately, the best upper bound on this function $f$ known to date
is exponential in $k$. The results reported in this paper would have
much simpler proofs if one could establish a polynomial upper bound
for the function $f$ in the previous theorem.

\wall{
Closely related to grids are \emph{walls}. Figure~\ref{fig:wall} b)
shows a $(6\times 6)$-wall.
An \emph{elementary $(k\times l)$-prewall} $W'$ is a graph with vertex
set $\{ (i,j) \st 1 \leq i\leq k, 1 \leq j \leq 2l\}$ in which two
vertices $(i,j)$ and $(i', j')$ are adjacent if, and only if, one of
the following possibilities holds: (1) $i'=i$ and $j' \in \{j-1,
j+1\}$ or (2) $j=j'$ and $i' = i+(-1)^{i+j}$. An \emph{elementary
  $(k\times l)$-wall} is obtained from an elementary $(k\times
l)$-prewall by removing the two vertices of degree $1$. An
\emph{$(k\times l)$-wall} $W$ is a graph isomorphic to a sub-division of
an elementary wall.

The cycles of minimal length in a wall $W$ are called \emph{bricks}.
 The \emph{height} $n$
of a wall is the number of rows of bricks and its \emph{width} $m$
the number of columns of bricks, i.e.~the height of an $(k\times
l)$-wall is $k-1$ and its width is $l-1$. 
A wall of \emph{order} $l$ is a wall of height and width $l$.
Finally, the \emph{nails} of a wall are the vertices of the
underlying elementary wall. Hence, in an elementary wall all vertices
are nails, in a general wall only some but including all vertices of
degree $3$. 

It is easily seen that any $G_{n,l}$ contains a wall of height $n$ and
width $l$ as
sub-graph and conversely any $(n,l)$-wall contains an $(n,l)$-grid as
minor (and this is the reason why width and height are defined as they
are). 
Hence the excluded grid theorem~\ref{thm:excl-grid} equally
applies to walls. However, as walls have maximum degree $3$,
Lemma~\ref{lem:top-minor} implies that if a wall occurs as a minor in
a graph, it occurs as a topological minor. For the purpose of this
paper, this makes it much easier
to work with walls than grids and we will therefore mostly work with
walls in the sequel.

}

\section{Monadic Second-Order Logic}
\label{sec:mso}

In this section we will introduce monadic second-order logic.
Intuitively, \emph{monadic second-order logic} is the extension of
first-order logic by quantification over sets of elements. That is, we
can use formulas of the form $\exists X \phi(X)$ which says that there
exists a set $X$ which satisfies the formula $\phi$. However, in the
context of graphs there are two natural options for what constitutes
an element: we can allow quantification over sets of vertices or
quantification over sets of edges. This leads to two different logics
which are sometimes referred to as $\MSO_1$ and $\MSO_2$,
respectively, where $\MSO_2$ allows quantification over sets of edges
and vertices whereas $\MSO_1$ only allows quantification over sets of vertices.
$\MSO_2$ is much more expressive than $\MSO_1$ as we can easily say
that a graph contains a simple path which contains every vertex, a
property that is not definable in $\MSO_1$. See below for an example
of a formula defining this property. 
For the purpose of this
paper it is convenient to introduce $\MSO_2$ as a logic on the
\emph{incidence representation of graphs}, which we will formally define below.

\subsection{Signatures and Structures. } 

We assume familiarity with basic notions of mathematical
logic (see~e.g.~\cite{EbbinghausFluTho94}).
A \emph{signature} $\sigma$
is a finite set of constant symbols $c\in\sigma$
and relation 
symbols $R\in \sigma$ where each relation symbol is equipped with its
arity $\ar(R) \in\N$. 

A \emph{$\sigma$-structure $\AA$} consists of a finite set
$A$, the \emph{universe} of $\AA$, an $r$-ary relation $R^\AA \subseteq
A^r$ for each
relation symbol $R\in\sigma$ of arity $r :=
\ar(R)$ and a
constant $c^\AA\in A$ for each constant symbol $c\in \sigma$.
We will denote structures by German letters $\AA, \BB, \DD$ and their
universes by corresponding Roman letters $A,B,D$.

In this paper we will only consider \emph{binary
  signatures}, where the maximal arity of relation symbols is $2$. An
example of classes of structures over binary signatures are the
skeletons of XML databases, i.e.~XML databases where the actual data
values are ignored. For instance, the database
\begin{center}
\begin{verbatim}
   <libraryholdings>
      <book>
         <author>YM</author>
         <title>EIAS</title>
      </book>
      <book>
         <author>RD</author>
         <title>GT</title>
      </book>
   </libraryholdings>
\end{verbatim}
\end{center}
can naturally be modelled as a structure $\DD$ over the signature
$\sigma := \{ \textit{book}, \textit{author}, \textit{title}
\}$, where the elements of the universe $D$ corresponds to the
individual tags \texttt{<libraryholdings>} etc. and edges represent
the child relation. 

Another natural interpretation of binary signatures is that structures
over these signatures are coloured graphs, i.e.~the binary relations
represent edges coloured by the relation name and the unary relations
represent vertex colours. 

To work with logics on graphs we have to specify how we want to represent
graphs as logical structures. 
Let $\sigmai := \{ V,
E, \in\}$ be a signature, where $V, E$ are unary and $\in$ is a
binary relation symbol. 
We can view a graph $G$ as a $\sigmai$-structure $\GG := \GG(G)$
with universe $\vrtG \dot\cup \edgeG$ and $V^\GG := \vrtG$, $E^\GG
:= \edgeG$ and $x\in^\GG e$ if $x\in \vrtG, e\in\edgeG$ and $x$ and
$e$ are incident in $G$. This is known as the \emph{incidence representation} of
graphs as opposed to the natural representation of graphs $G$ as structures
$\GG$ over the signature $\{E\}$, where the universe is $\vrtG$ and
$E^\GG := \edgeG$. 

We now extend the definition of tree-width from graphs to arbitrary
relational structures.

\begin{defi}\label{def:struct-tw}
  Let $\sigma$ be an at most binary signature. The \emph{tree-width
    $\tw(\AA)$} of a $\sigma$-structure $\AA$ is defined as the
  tree-width $\tw(\GGG(\AA))$ of its Gaifman-graph (see Definition~\ref{def:gaifman}).
\end{defi}

In the context of graphs and tree-width it might be worth noting that
the tree-width of a graph is the same as the tree-width of its
standard or incidence representation.

\begin{defi}\label{def:signatures}
  For the rest of this paper we fix a signature $\sigmacol := \{ V, E,
  \in, B, R, C_0, C_1\}$, where $\in$ is a binary relation symbol and $V, E,
  B, R, C_0, C_1$ are unary.

  We define the signature $\sigmaw := \{ V, E, \in, C_0,
  C_1\}$ and $\sigmaord := \sigmaw \cup \{\leq\}$.
\end{defi}

\subsection{Definition of Monadic Second-Order Logic}

The class of formulas of \emph{monadic
second-order logic}
over a signature $\sigma$, denoted $\MSO[\sigma]$,
is defined as the extension of first-order logic
by quantification over sets of elements.
That is, in addition to first-order
variables, which we will denote by small letters
$x, y, ...$, there are unary, or monadic, second-order variables $X, Y, ...$
ranging over sets of elements. Formulas of
$\MSO[\sigma]$ are then built up inductively by
the rules for first-order logic with the following
additional rules: if $X$ is a monadic second-order
variable and $\phi\in\MSO[\sigma
\dot\cup \{X\}]$, then $\exists
X\phi\in\MSO[\sigma]$ and $\forall X
\phi\in\MSO[\sigma]$ with the obvious semantics
where, e.g., a formula $\exists X\phi$ is true in a
$\sigma$-structure $\AA$ with universe $A$ if there is a subset
$U'\subseteq A$ such that $\phi$ is true in
$\AA$ if the variable $X$ is interpreted by $U'$. We
denote this by $(\AA, U')\models \phi(X)$. If
$\phi(x)$ is a formula with a free first-order variable $x$,
$\AA$ is a structure and $a\in A$, we write $\AA\models \phi[v]$, or
$(\AA, v)\models \phi$, to say that $\phi$ is true in $\AA$ if $x$ is
interpreted by $a$. We write $\phi(\AA)$ for the set
$\{ v\in A \st \AA\models\phi[v]\}$.

As explained above, when viewed as a logic on graphs, the expressive
power of monadic second-order logic depends on whether a
graph is represented by its standard representation or by its
incidence representation. It has become common terminology to refer to
$\MSO$ on graphs represented by their standard representation as
$\MSO_1$ and to use $\MSO_2$ to indicate that graphs are represented
by their incidence structures.

We will follow this terminology. Therefore, if $\sigma$ is an at most
binary signature, we define $\MSO_2[\sigma]$ to be monadic second-order logic over the signature
$\sigma\dot\cup \{ V, E, \in\}$ where $\sigma$-structures are
represented as incidence structures in the obvious way.
The main theorem stated in the introduction can therefore equivalently
be stated as a theorem on structures over a signature $\tau := \{ V,
E, \in, R_1, \dots, R_k, U_1, \dots, U_l, c_1, \dots, c_s\}$
containing at least two binary relation symbols $R_i$ and two unary relation
symbols $U_i$.

In this paper we will almost exclusively use the incidence representation
and therefore agree that $\MSO$ always refers to $\MSO_2$ unless
explicitly stated otherwise. Also, we will always make the signatures
we work with precise to avoid confusion.

We will not distinguish
notationally between a graph $G$ and its incidence representation $\GG$ and
will simply write $G$. To simplify the presentation of formulas, we
agree on the following notation.

\begin{Notation}
  We will write $\exists X\subseteq V\phi$ and $\exists F\subseteq
  E\phi$ as abbreviation for $\exists X\big((\forall x x\in X \rightarrow
  x\in V) \wedge \phi\big)$ and $\exists F\big((\forall x x\in F \rightarrow
  x\in E) \wedge \phi\big)$ to indicate that $X$ is a set of vertices and
  $F$ is a set of edges. $\forall X\subseteq V$ and $\forall
  F\subseteq E$ are defined analogously. 

  We write $e\cap X\not=\emptyset$
  for $\exists u\in V(u\in e \wedge u\in X)$ and similarly $e\subseteq
  X$ for $\forall u(u\in e\rightarrow u\in X)$ to say that $X$ contains
  an endpoint (both endpoints, resp.) of $e$. 
  Also, we will use
  notation such as $X\cap Y\not=\emptyset$, $X\subseteq Y$, ... with
  the obvious meaning.

  We will often use set variables $P$ which are intended to contain
  the edges of a path in a graph. The following notation helps to
  simplify formulas speaking about paths. 
  If $P$ is a variable denoting a set of edges then we write $x\in V(P)$ for the formula
  $\exists e (e\in P \wedge x\in e)$ expressing that $x$ occurs as an endpoint
  of an edge $e$ in $P$. Furthermore, we write $\{x, y\} \in P$ for the
  formula $x\not=y \wedge \exists e \in P (x\in e \wedge y\in e)$
  saying that $\{ x, y\}$ is an edge in $P$.

  Finally, we write
  $\exists^{\leq 2} x \phi$ for the formula
  $\exists x \exists y \big(\phi(x) \wedge \phi(y) \wedge \neg
  \exists z (z\not= x \wedge 
  z\not= y \wedge \phi(z)\big)$ expressing that there are at most two
  vertices satisfying $\phi$. We will also use $\exists^{=1},
  \exists^{\leq 1}$ with the obvious meaning.
\end{Notation}

\subsection{Examples} 

\begin{exa}\label{ex:3col}
  To give an example consider the following
  $\MSO$-formula $\phi$ over the signature $\sigmai$.
  \[
     \exists C_1, C_2, 
     C_3 \subseteq V \Big(\forall x\in V \bigvee_{i=1}^3
     x\in C_i \wedge \forall e \in E \bigwedge_{1\leq i\leq 3} \neg
     e\subseteq C_i\Big), 
  \]
  where $x,y$ are first-order variables and $C_1, C_2, C_3$ are
  second-order variables. The formula expresses in a
  $\sigmai$-structure $G$ that there are three sets of vertices so that
  every vertex occurs in at least one of the sets but no edge has
  both endpoints in the same set. Hence, $G\models \phi$ if, and
  only if, $G$ is $3$-colourable.\ExQed
\end{exa}

\begin{exa}
  As a second example we define a formula $\phi_{\textit{Ham}}$ true in a graph $G$
  if, and only if, the graph contains a Hamiltonian path, i.e. a
  simple path containing every vertex. 

  The formula $\textit{conn}(P)$ defined as
  \[
     \textit{conn}(P) := \forall X\subseteq V\big[\big(
     \begin{array}{l}
       V(P) \cap X \not= \emptyset\ \wedge\\
     \forall e\in P(e\cap X \not= \emptyset \rightarrow e\subseteq X)
   \end{array}
   \big) \rightarrow V(P) \subseteq X\big]
  \]
  says that if $X$ is  any set of vertices containing a vertex $x\in V(P)$
  which is closed under edges $e\in P$, i.e.~if one endpoint of $e$
  is in $X$ then both are, then $X$ must contain all vertices of
  $V(P)$. Clearly, this formula can only be true of a set $P$ of
  edges if $P$ induces a connected sub-graph. 
  The next formula expresses that $P$ induces an
  acyclic graph.
  \[
   \textit{ac}(P) := \neg \exists s,t \in V \exists P, P'\subseteq E
   \big(s\not= t \wedge \textit{conn}(P) \wedge \textit{conn}(P') \wedge V(P)\cap V(P') =
   \{ s, t\}\big)
  \]
  The formula states that there are no two distinct vertices $s$ and $t$ and
  two connected sub-graphs $P$ and $P'$ such that $s$ and $t$ are
  contained both in $P$ and $P'$ but otherwise $P$ and $P'$ are vertex
  disjoint. Clearly, any cyclic graph contains such $s, t, P, P'$ but
  no acyclic graph does.
  
  Hence,
  $\textit{conn}(P)\wedge\textit{ac}(P)$ says that
  $P$ induces a tree. Now the formula
  \[
     \textit{path}(P) := \textit{ac}(P) \wedge\textit{conn}(P) \wedge
     \forall x\in V
   \exists^{\leq 2} e(e\in P \wedge x \in e) 
  \]
  says that $P$ is a tree and every vertex has degree at most $2$
  in the graph induced by $P$. Hence, $P$ is a path. Finally,
  \[
  \phi_\textit{Ham} := \exists P \textit{path}(P) \wedge V \subseteq V(P)
  \]
  expresses that the graph contains a Hamiltonian path. Here we
  crucially need quantification over sets of edges (which is implicit
  in the incidence encoding of graphs) as the
  Hamiltonian-path property is not expressible in \textup{MSO} without edge
  set quantification.\ExQed
\end{exa}

\begin{exa}\label{ex:grid}
  We now give a much more substantial example which will be used in
  the  proof of the main results of this paper. In particular, we will
  show that grids can be defined in monadic second-order logic. 

  We first establish the following characterisation of grids which can
  then easily be turned into an $\MSO$-formulation.

  Let $G$ be a graph and $\Hp,\Vp$ be two sets of pairwise vertex
  disjoint paths, which we think of the horizontal and vertical paths
  in the grid. Then $\Hp \cup \Vp$ is a grid if, and only
  if, the following conditions are true.
  \begin{enumerate}[(1)]
    \item Any two $P\in \Vp, Q\in\Hp$ intersect in exactly one vertex
      and every vertex of the graph is contained in the intersection
      of two such paths $P\in \Vp, Q\in\Hp$.
    \item There are distinct $L, R\subseteq \Vp$, the left-most and right-most
      path of  the grid, such that every $Q\in \Hp$ intersects $L$
      and $R$ in one endpoint.
      Analogously, there are distinct $T, B\subseteq \Vp$, the upper-most and lower-most
      path of the grid, such that every $P\in \Hp$ intersects $T$
      and $B$ in one endpoint.
    \item Let us define an order $\leq_P$ on the vertex set of a path $P\in\Vp$
      such that $x \leq_P y$, for $x,y\in V(P)$ if $x$ is closer to
      the endpoint of $P$ in $T$ than $y$, i.e.~if the unique path
      from $y$ to the endpoint of $P$ in $T$ also contains $x$. We
      write $x <_P y$ for the corresponding strict order.  Analogously,
      we define $x \leq_Q y$, for $Q\in\Hp$ and $x,y\in V(Q)$, if $x$
      is closer to endpoint of $Q$ in $L$, i.e.~the unique path from $y$ to the
      endpoint of $Q$ in $L$ contains $x$. Again $x<_Q y$ denotes the
      strict variant.

      Let $P, P' \in \Vp$ and $Q, Q'\in \Hp$ and let $x \in V(P\cap
      Q)$, $x'\in V(P'\cap Q)$ and $y\in V(P \cap Q')$ and $y' \in
      V(P'\cap Q')$. Then,
      \begin{iteMize}{$\bullet$}
      \item 
        $x <_P y$ if, and only if, $x' <_{P'} y'$ and
      \item $x<_Q x'$ if, and only if, $y <_{Q'} y'$.
      \end{iteMize}

      That is, we require that all ``horizontal'' paths $Q, Q'\in \Hp$
      cross all vertical paths $P, P'\in\Vp$ in the same order, seen
      from the top, and that all ``vertical'' paths $P,P'\in\Vp$
      cross all horizontal paths $Q,Q'\in\Hp$ in the same order seen
      from the ``left''.
  \end{enumerate}

  \noindent It is easily seen that if $\Hp$ is the set of horizontal paths and
  $\Vp$ the set of vertical paths in a grid, then $\Vp, \Hp$ satisfy
  these conditions. Conversely, let $\Vp,\Hp$ be two sets of pairwise
  disjoint paths satisfying conditions $1$ to $3$ then the graph
  induced by $\Vp,\Hp$ is a grid. 

  We show next how these conditions can be formalised by a formula
  $\phi_{\textit{grid}}(\Hp, \Vp)$. To simplify the presentation, we
  will use second-order variables $\PPP, \QQQ, \Vp, \Hp$ which we will
  always ensure to be interpreted by sets of pairwise disjoint paths.

  We first define some basic formulas which will be used frequently
  later on. 

  The formula 
  \[
     \sodp(\PPP) := \PPP \subseteq E \wedge \textit{ac}(\PPP) \wedge
     \forall x \exists^{\leq 2} e\in \PPP (x\in e) 
  \]
  expresses that $\PPP$ is a set of edges inducing an acyclic sub-graph in which every
  vertex has degree at most $2$. Hence $\PPP$ must be a set of
  pairwise vertex disjoint paths.

  The next formula
  \[
     \maxpath(P, \PPP) := \PPP \subseteq E \wedge P\subseteq \PPP
     \wedge \textit{path}(P) \wedge \forall P' \big(P \subseteq P' \wedge
     P'\subseteq \PPP \rightarrow \neg \textit{path}(P')\big)
  \]
  states that $P$ is a maximal path in $\PPP$, hence it one of the
  paths in the set $\PPP$ of pairwise disjoint paths. We will write
  $\exists P\in \PPP$ as abbreviation for $\exists P \maxpath(P,
  \PPP)$ and likewise for $\forall P\in \PPP$.

  Finally, 
  \[
  \phiendpoint(x, P) := \pAth(P) \wedge \exists^{=1} e \in P(x\in e)
  \]
  defines that $x$ is an endpoint of the path $P$.

  Now the conditions above can easily be defined in $\MSO$ as follows.
  The formula $\phi_0 := \sodp(\Vp) \wedge \sodp(\Hp)$ ensures that
  $\Vp$ and $\Hp$ are interpreted by sets of pairwise vertex disjoint
  paths.

  The formula
  \[
    \phi_1(\Vp, \Hp) :=
    \begin{array}{l}
      \forall P\in \Vp \forall Q\in\Hp \exists^{=1} x
      \in V (x \in V(P) \cap V(Q))\ \wedge\\
      \forall x\in V \exists^{=1} P
      \exists^{=1} Q \big( x\in V(P) \wedge x \in V(Q)\big)
    \end{array}
  \]
  expresses the first condition above.

  The formula $\phi_2(\Vp, \Hp, L, R, T, B)$
  \[
   \phi_2  :=
    \begin{array}{l}
      L\in \Vp \wedge R \in \Vp \wedge T \in \Hp \wedge B \in \Hp\ \wedge\\
      \forall P \in \Vp \exists x_1, x_2 \in V(P)
      \big(\phiendpoint(x_1, P) \wedge \phiendpoint(x_2, P) \wedge x_1
      \in V(T) \wedge x_2 \in V(B)\big)\ \wedge\\
      \forall Q \in \Hp \exists x_1, x_2 \in V(Q)
      \big(\phiendpoint(x_1, Q) \wedge \phiendpoint(x_2, Q) \wedge x_1
      \in V(L) \wedge x_2 \in V(R)
    \end{array}
  \]
  expresses $L, R, T, B$ satisfy the requirements outlined in
  Condition 2.

  Finally, we define a formula expressing Condition $3$. The formula
  \[
     \phi_P(x, y, P) :=
     \begin{array}{l}
       \exists p \in V(P) \cap V(T)\ \wedge\\
       \forall P'\big( P' \subseteq P \wedge \pAth(P') \wedge p
     \in V(P') \wedge y\in V(P') \rightarrow x \in V(P')\big)
   \end{array}
  \]
  defines the ordering $\leq_P$ on a path $P\in \Vp$. It states that
  $x \leq_P y$ if every sub-path of $P$ containing the endpoint in $T$
  and $y$ also contains $x$. Analogously, the formula
  \[
     \phi_Q(x, y, Q) :=
     \begin{array}{l}
       \exists p \in V(Q) \cap V(L)\ \wedge\\
       \forall Q'\big( Q' \subseteq Q \wedge \pAth(Q') \wedge p
     \in V(Q') \wedge y\in V(Q') \rightarrow x \in V(Q')\big)
   \end{array}
  \]
  defines the ordering $\leq_Q$ on a path $P\in \Hp$.

  Hence, the formula
  \[
     \phi_3 := \forall P, P'\subseteq \Vp\ \forall Q, Q'\subseteq \Hp
     \left(
     \begin{array}{l}
       \exists x \in V(P\cap Q) \exists x' \in V(P'\cap Q)\\
       \exists y \in V(P\cap Q') \exists y' \in V(P'\cap Q')\\
       \big(\phi_P(x, y, P) \leftrightarrow \phi_P(x', y', P')\big)
       \ \wedge\\
       \big(\phi_Q(x, x', Q) \leftrightarrow \phi_Q(y, y', Q')\big)
     \end{array}
     \right)
  \]
  defines Condition $3$.
  
  Taken together, the formula
  \[
     \phi_{grid-border}(L, R, T, B, \Vp, \Hp) := \bigwedge_{i=0}^3 \phi_i
  \]
  expresses that $\Vp, \Hp$ form a grid with borders $L, T, R, B$,
  clock-wise from left. The formula $\phi_{\textit{grid}}$ can
  therefore be defined as $\exists L, T, R, B \subseteq E\
  \phi_{\textit{grid-border}}(L, R, T, B, \Vp, \Hp)$. \ExQed
\end{exa}

As the examples show, once we have established a
few basic formulas such as \textit{path} and
\textit{ac}, many properties of graphs can very
easily be expressed in $\MSO$. We are therefore
particularly interested in the problem of deciding
whether a given $\MSO$-formula is true in a graph
$G$.

\subsection{$\MSO$-Transductions}
\label{sec:mso-inter}

A useful tool in the proof of our main results in this paper is the
concept of logical transduction, which for our purposes play a similar role
to many-one reductions in complexity theory. 
Essentially, a transduction is a way of defining one logical structure
inside another. This concept is usually referred to as
\emph{interpretations} in model theory, see e.g.~\cite{Hodges97}
for details. However, we will use interpretations in the ``wrong''
direction and therefore follow Courcelle's notation and call them
transductions (see e.g.~\cite{CourcelleOumO07}). 

\begin{defi}
   Let $\sigma$ and $\tau$ be signatures and let $\tplX$ be a
   tuple of monadic second-order variables. An  
   \emph{\MSO-transduction of $\sigma$ to $\tau$ with parameters
   $\tplX$} is a tuple $\Theta := \big(\phivalid,$ $\phiuniv(x)$,
   $\phi_\sim(x,y)$, $(\phi_R(\tplx))_{R\in\tau}\big)$ of
   $\MSO[\sigma \dot\cup \tplX]$-formulas, where the arity of
   $\tplx$ in   $\phi_R(\tplx)$ is $\ar(R)$, such that for all
   $\sigma$-structures $\AA$ and interpretations $\tplY\subseteq
   A$ of $\tplX$ with $(\AA, \tplY)\models
   \phivalid$, $\phi_\sim$ defines an equivalence
   relation on $\phiuniv(\AA)$ and if $R\in\tau$ or arity $r$ and
   $\tpla := a_1, \dots, a_r, \tplb := b_1, \dots, b_r \in A^r$ are
   tuples such that $\AA \models \phi_\sim(a_i, b_i)$ for all $i$ then
   $\AA\models \phi_R(\tpla)$ if, and only if, $\AA\models
   \phi_R(\tplb)$. 
\end{defi}

With any transduction $\Theta$ we associate a map taking a
$\sigma$-structure $\AA$ and $\tplY\subseteq A$ such that
$(\AA, \tplY)\models \phivalid$
to a $\tau$-structure $\BB$ with universe $B :=
\phiuniv(\AA,\tplY)_{/\phi_\sim(\AA,\tplY)} := \{ [v]_\sim \st
(\AA,\tplY)\models \phiuniv(v)\}$ where $[v]_{\sim}$ denotes the
equivalence
class of $v$ under the equivalence relation defined by
$\phi_\sim(A,\tplY)$. 
For $R\in\tau$ of arity
$r := \ar(R)$ we define $R^\BB := \{ ([a_1], \dots, [a_r]) \st
(\AA,\tplY)\models \phi_R(a_1, \dots, a_r)\}$.

For any $\sigma$-structure $\AA$ we define 
\[
  \Theta(\AA) := \{ \Theta(\AA,
    \tplY) \st \tplY \subseteq A, (\AA,\tplY)\models \phivalid \}.
\]
If $\CCC$ is a class of $\sigma$-structures then 
\[
   \Theta(\CCC) := \bigcup \{ \Theta(\AA) \st \AA\in\CCC\}.
\]

Furthermore, any interpretation $\Theta$ also defines a
translation of
$\MSO[\tau]$-formulas $\phi$ to $\MSO[\sigma]$-formulas
$\phi'$ by replacing occurrences of relations $R\in\tau$
by their defining formulas $\phi_R\in\Theta$ in the usual
way (see \cite{Hodges97} for details). For notational convenience we
define $\Theta(\phi) := \phivalid \wedge \phi'$. The following lemma
is then easily proved.

\begin{lem}
   Let $\Theta$ be an $\MSO$-transduction of $\sigma$ in $\tau$ with parameters
   $\tplX$. For any
   $\sigma$-structure $\AA$ and assignment $\tplY \subseteq
   A$ to $\tplX$ such that $(\AA, \tplY)\models\phivalid$ and any $\MSO[\tau]$-formula $\phi$ we have
   $\Theta(\AA, \tplY) \models \phi$ if, and only if,
   $(\AA, \tplY)\models\Theta(\phi).$
\end{lem}

We will be using the previous lemma as
summarised in the next corollary.

\begin{cor}\label{cor:inter}
  Let $\Theta := (\phivalid, ...)$ be an $\MSO$-transduction of $\sigma$ in $\tau$ with
  parameters $\tplX$. For any $\sigma$-structure $\AA$ and
  $\MSO[\tau]$-formula $\phi$ we have
  \[
     \AA \models \exists \tplX (\phivalid \wedge \Theta(\phi)) \text{
       if, and only if, there exists } \BB\in\Theta(\AA) \text{ s.t. }
     \BB\models \phi. 
  \]
\end{cor}

\begin{exa}
  We will define a transduction  
  $\Theta := (\phivalid, \phiuniv, \phi_\sim, \phi_V, \phi_E,
  \phi_\in)$ with parameters $\PPP, \QQQ$
  from $\sigmai$ to $\sigmai$ so that for any $\sigmai$-structure
  $\AA$ and two sets $\PPP, \QQQ$ of disjoint paths in $\AA$, $\Theta(\AA,
  \PPP, \QQQ)$ is the incidence representation of the intersection graph
  $\III(\PPP, \QQQ)$ (see Definition~\ref{def:intersection-graph}). 

  The formula $\phivalid$ is simply defined as 
  \[
    \phivalid(\PPP, \QQQ) := \sodp(\PPP) \wedge \sodp(\QQQ)
  \]
  stating that $\PPP$ and $\QQQ$ are
  sets of pairwise disjoint paths (see Example~\ref{ex:grid} for the
  formula $\sodp$). 

  To define the $\phi_V, \phi_E$, recall that the vertices of $\III := \III(\PPP, \QQQ)$ are the paths
  in $\PPP\dot\cup \QQQ$ and that two vertices $P, Q$ are adjacent if the
  paths intersect. 
  We will represent a path $P\in \PPP\dot\cup \QQQ$, and hence the
  corresponding vertex in $\III$, by the set of
  edges of $\AA$ occurring only in $P$ and in no other path in
  $\PPP\dot\cup\QQQ$. Note that as $\PPP$ and $\QQQ$ are sets of
  pairwise disjoint paths, such edges must always exist, whereas it
  could happen that every vertex of $P$ also occurs as a vertex of
  another path.

  Towards this goal, the formula 
  \[
    \uniqueedge(x, P) := x \in P \wedge \forall Q \big( P\not= Q
    \wedge (\maxpath(Q, \PPP) \vee \maxpath(Q, \QQQ)) \rightarrow \neg x \in Q\big),
  \]
  where $P\not= Q$ is an abbreviation for $\exists x \in V(P)
  \setminus V(Q)$ saying that $P$ and $Q$ are not the same path,
  states that $x$ is an edge unique to $P$.

  The formula 
  \[
    \phiuniv^V(x; \PPP, \QQQ) :=
    \begin{array}{l}
      \exists P  \big(\textit{maxpath}(P, \PPP) \wedge \uniqueedge(x,
      P)\big)\ \vee\\
      \exists Q \big(\textit{maxpath}(Q, \QQQ) \wedge \uniqueedge(x, Q)\big)
  \end{array}
  \]
  defines the set of edges unique to a path in $P\dot\cup Q$. Correspondingly, the formula
  \[
     \phi_\sim^V(x,y;\PPP,\QQQ) := 
     \begin{array}{l}
      \exists P  \big(\textit{maxpath}(P, \PPP) \wedge \uniqueedge(x,
      P) \wedge \uniqueedge(y,
      P)\big)\ \vee\\
      \exists Q \big(\textit{maxpath}(Q, \QQQ) \wedge \uniqueedge(x,
      Q) \wedge \uniqueedge(y,  Q)\big)
     \end{array}
  \]
  defines two vertices of $\III$ to be equivalent if the are unique
  edges of the same path in $\PPP\dot\cup\QQQ$.
  
  To define the edges of $\III$, we will represent an edge $\{ P, Q\}$ in $\III$ by the
  set of vertices in $V(P \cap Q)$.
  The formula 
  \[
    \phiuniv^E(x; \PPP, \QQQ) := x \in V \wedge \exists P \exists Q
    \big(\textit{maxpath}(P, \PPP) \wedge \textit{maxpath}(Q, \QQQ) \wedge
    x \in V(Q \cap P)\big)
  \]
  defines the set of all vertices which occur in the intersection of
  two paths. Correspondingly, the formula $\phi_\sim^E(x, y; \PPP,
  \QQQ)$ defined as 
  \[
      \exists P \exists Q
     \Big(\textit{maxpath}(P, \PPP) \wedge \textit{maxpath}(Q, \QQQ) \wedge
      x \in V(Q \cap P) \wedge y \in V(Q \cap P)\Big)
  \]
  defines two vertices to be equivalent if they occur together
  in the intersection of the same two paths.
  
  Hence, the vertex set of the incidence representation of $\III$ is
  defined by $\phiuniv(x; \PPP, \QQQ) := \phiuniv^V \vee \phiuniv^E$
  and $\phi_\sim(x, y; \PPP, \QQQ) := \phi_\sim^V \vee \phi_\sim^V$
  and the relations $E$ and $V$ are defined  by 
  $\phi_E(x;\PPP,\QQQ) := \phiuniv^E$ and $\phi_V(x; \PPP, \QQQ) :=
  \phiuniv^V$.  

  All that remains is to define the formula $\phi_\in(x, y, \PPP,
  \QQQ)$. But this can easily be done as a vertex $x$ in $\III$, say
  corresponding to a path $P\in\PPP$ and therefore represented by the
  set of unique edges of $P$, is incident to an edge $e$ of $\III$,
  corresponding to the intersection of two paths $P'\in\PPP$ and $Q\in\QQQ$ and
  thus represented by the set $V(P') \cap V(Q)$, if $V(P') \cap V(Q')
  \subseteq V(P)$, i.e.~if $e\in V(P)$.
  This is expressed by the following formula
  \[
    \phi_\in(x, e, \PPP, \QQQ) :=
    \begin{array}{l}
      \phi_V(x) \wedge \phi_E(e)\ \wedge\\
    \exists P \big((\maxpath(P, \PPP) \vee \maxpath(P, \QQQ)) \wedge
    \uniqueedge(x, P) \wedge e \in P\big).

  \end{array}
  \]
  This completes the transduction $\Theta$. \ExQed
\end{exa}

\section{The Complexity of Monadic Second-Order Logic}
\label{sec:complexity}

The \emph{model checking problem} $\MC(\MSO)$ for $\MSO$ is defined as the
problem, given a structure $G$ and a formula $\phi\in\MSO$, to decide
if $G\models \phi$. By a reduction from the \PSPACE-complete 
\emph{Quantified Boolean Formula Problem} (QBF) -- the problem to decide whether a quantified Boolean formula is
true -- we easily get that 
$\MC(\MSO)$ is \PSPACE-hard (see \cite{Vardi82}). In fact, the problem
is \PSPACE-complete as membership in \PSPACE is easily seen.

However, the hardness result crucially uses the fact that the
formula is part of the input (and in fact holds on
a fixed two-element structure), whereas we are
primarily interested in the complexity of checking
a fixed formula expressing a graph property in a
given input graph. We therefore study model-checking problems in
the framework of \emph{parameterized complexity} (see \cite{FlumGro06}
for background on parameterized complexity).

\begin{defi}
  Let $\CCC$ be a class of $\sigma$-structures. The
  \emph{parameterized model-checking problem} $\pMC(\MSO, \CCC)$ for
  $\MSO$ on $\CCC$ is defined as the problem to decide, given $G\in \CCC$ and
  $\phi\in\MSO[\sigma]$, if $G\models \phi$. The \emph{parameter} is
  $|\phi|$.

  $\pMC(\MSO, \CCC)$ is \emph{fixed-parameter tractable} (fpt), if
  there exists a computable function $f\st\N\rightarrow \N$ and a
  $c\in\N$ such that for all
  $G\in\CCC$ and $\phi\in\MSO[\sigma]$, $G\models \phi$ can be
  decided in time $f(|\phi|)\cdot |G|^c$. The problem is in the class XP, if it can be decided
  in time $|G|^{f(|\phi|)}$. 
\end{defi}

In Example~\ref{ex:3col} we have seen that the $\NP$-complete 
$3$-Colourability problem is definable in $\MSO$. Hence,
$\MC(\MSO, \textsc{Graphs})$, the 
model-checking problem for $\MSO$ on the class of all graphs, is not
fixed-parameter tractable unless $P=\NP$. 
However, Courcelle~\cite{Courcelle90} proved that if we restrict
the class of admissible input graphs, then we can
obtain much better results. 
Recall the definition of tree-width of structures from
Definition~\ref{def:struct-tw}.

\begin{thm}[\cite{Courcelle90}] \label{theo:courcelle}
  There is an algorithm which, given a graph $G$ in its incidence
  representation and an $\MSO$-formula $\phi$, decides ``$G\models
  \phi$?'' in time $f(|\phi|+\tw(G))\cdot |G|$. 

  Hence, $\MC(\MSO, \CCC)$ is fixed-parameter tractable on any class $\CCC$
  of structures of tree-width bounded by a constant.  
\end{thm}

Courcelle's theorem gives a sufficient condition
for $\MC(\MSO, \CCC)$ to be tractable. 
The obvious counterpart are sufficient conditions for intractability,
i.e.~what makes $\MSO$-model checking hard?
Garey, Johnson and
Stockmeyer \cite{GareyJohSto74} proved that  3-Colourability remains
\NP-hard on the class of planar graphs of degree at most $4$.
It follows that unless $P=\NP$, $\MC(\MSO, \textsc{Planar})$ is not
fixed-parameter tractable, where \textsc{Planar} denotes the class of
planar graphs. 
However, this result only indirectly relates tractability of
$\MSO$ model-checking on a class $\CCC$ to its
tree-width.
It would therefore be interesting to investigate
whether Courcelle's theorem can be extended to class of
unbounded tree-width or conversely, which bounds
on the tree-width of a class $\CCC$ 
prohibit tractable $\MSO$-model-checking.
As we have seen above, large tree-width of graphs implies the
existence of large grid-minors and it is well-known that
$\MSO$-model checking is hard on the class of grids. We will make use of this fact below and therefore repeat
the statement here.

\begin{defi}
  Recall from Definition~\ref{def:signatures} the signature $\sigmaw :=
  \{ V, E, \in, C_0, C_1\}$ of coloured grids, where $V, E, C_0, C_1$ are unary relation
  symbols and $\in$ is a binary relation symbol. A $\sigmaw$-structure
  $G$ is a \emph{coloured $l\times l$-grid} if its
  $\sigmai$-reduct $W_{|\{V, E, \in\}}$ is an $l\times
  l$-grid. 

  $G$ \emph{encodes a word
  $w := w_1\dots w_n\in\Sigma^n$ with power $d$} if $l\geq n^d$, and $C_0 \cap C_1 = \emptyset$ and if
  $\{ v_{1, i} \st 1\leq i\leq l\}$ are the vertices on the bottom row
  then $v_{1, i}\in C_0$ if $w_i = 0$ and $v_{1, i}\in C_1$ if $w_i =
  1$, for all $1\leq i\leq n$. 
\end{defi}

The following theorem is a well-known fact about the complexity of $\MSO$.

\begin{thm}\label{lem:grids}
  For $d\geq 2$ let $\GG_d$ be the class of coloured grids encoding
  words with power $d$. Then $\MC(\MSO,\GG_d)$ is not in XP
  unless $P=\NP$. 
\end{thm}

The theorem follows immediately from the following
lemma, whose proof is standard. 

\begin{lem}\label{lem:TM-enc}
  Let $M$ be a non-deterministic $n^d$-time bounded Turing-machine.
  There is a formula $\phi_M\in\MSO$ such that for all words
  $w\in\Sigma^\star$, if $G$ is a coloured grid encoding
  $w$ with power $d$, then $W\models \phi_M$ if, and only if, $M$
  accepts $w$.  Furthermore, the formula $\phi_M$ can be constructed
  effectively from $M$.  The same holds if $M$ is an alternating
  Turing-machine with a bounded number of alternations, as they are
  used to define the polynomial-time hierarchy.
\end{lem}

\begin{figure}[ht]
  \centering
  \begin{tabular}{cc}
    \includegraphics[width=7cm]{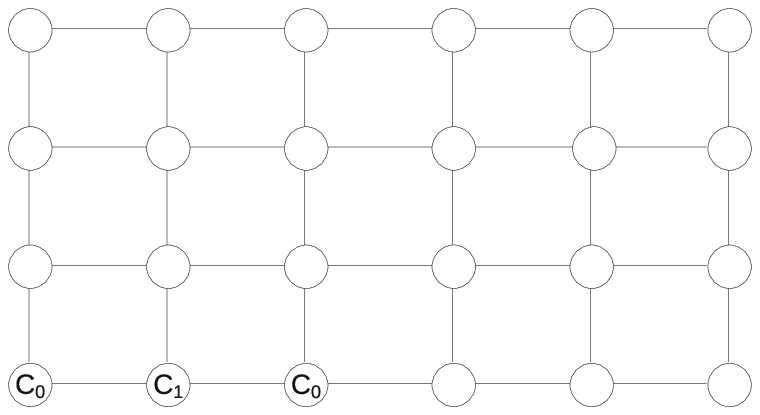} &
    \includegraphics[width=7cm]{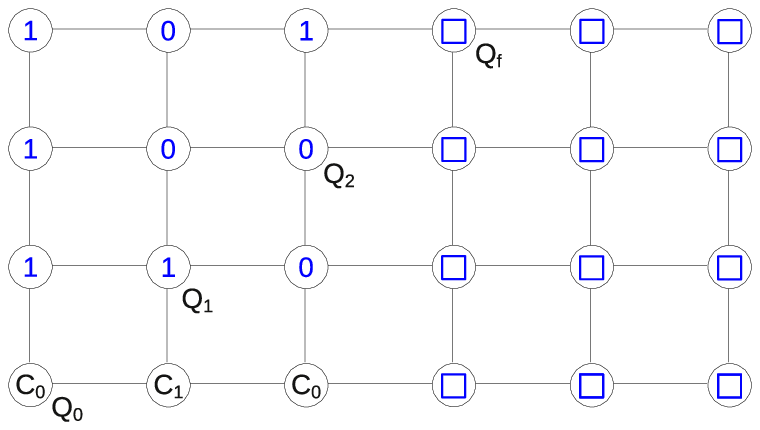} \\
    Coloured grid encoding $010$ &
    Quantifiers guessing a run of $M$
  \end{tabular}
  \caption{Guessing a run of a Turing-machine in $\MSO$.}
  \label{fig:TM}
\end{figure}

\begin{Proof}[Proof sketch]
  The main idea of the proof is to use existential set quantification
  and the grid to guess the time-space diagram of a successful run $R$
  of the Turing-machine $M$ on input $w$. Figure~\ref{fig:TM}
  illustrates this idea.

  The grid on the left hand side encodes the word $010$ through the
  three vertices in $C_0$ and $C_1$. We can then use existentially
  quantified monadic second-order variables $Q_0, Q_1, Q_2, Q_f$,
  $S_0, S_1, S_\Box$ so that $Q_s$ contains a vertex $(i, j)$ if the
  Turing machine $M$ would be in state $q_s$ after $i$ steps in the
  run $R$ with the read/write head scanning position $j$. A vertex
  $(i, j)$ appears in $S_0$ if after $i$ steps the tape cell $j$
  contains symbol $0$, and likewise for $S_1, S_\Box$ denoting cells
  containing $1$ and the blank symbol $\Box$.

  That these existentially quantified variables indeed encode a valid
  and accepting run of $M$ on input $w$ can easily be formalised in
  first-order logic, as the content of position $(i, j)$ only depends
  on the content of $(i-1, j-1), (i-1,j), (i-1, j+1)$ and hence is a
  local property.

  The reason we use a grid encoding a word with \emph{power $d$} is
  that we need the grid to be large enough so that we can guess the
  complete run of the machine $M$ on input $w$, and if $M$ is $n^d$
  time bounded, then it can use up to $n^d$ steps and $n^d$ tape cells.
\end{Proof}

\section{A High Level Description of the Main Proof}
\label{sec:overview}

In this section we give a high level description of the proof of the
main theorem~\ref{theo:main}. We want to show that if $\CCC$ is a
class of $\sigma$-structures whose tree-width is $(\log^c n,
p)$-unbounded, for some large enough $c$ and polynomial $p$, and which
satisfies the conditions of the theorem, then model-checking for
$\MSO$ is not in XP on the class $\CCC$.
At the core of the proof is a reduction from $\MSO$ model-checking on
$\CCC$ to model-checking of $\MSO$ on the class of coloured grids
which we have already seen to be intractable. 
We now present a first idea of how to do this. The idea will not work but it
helps to illustrate how the theorem is actually proved.

We show intractability of
$\pMC(\MSO, \CCC)$ by reducing an $\NP$-complete problem $P$ to
$\pMC(\MSO, \CCC)$ as follows. 
Given a word $w$ of length $n$, we choose a graph $G\in\CCC$ of
large enough tree-width. By the excluded grid
theorem~\ref{thm:excl-grid}, $G$ contains a large grid minor. Such a
grid-minor can be defined in monadic second-order logic: we have
already seen how to say that a graph is a grid, all we need to do is
to extend this to say that a graph contains a grid-minor. This
requires some work, but can be done. As $\CCC$ is closed under
colourings, we can use vertex colours to encode the word
$w$ in this grid-minor as indicated in the previous
section. Hence, given $w$ we have constructed a graph $G\in\CCC$ of large
enough tree-width and from this get a graph $G_w$ with a large
grid-minor encoding the word $w$. Furthermore, this grid-minor
encoding $w$ can be defined by $\MSO$-formulas, more precisely there
is an $\MSO$-transduction taking the graph $G_w$ and mapping it to the
coloured grid $H$ encoding $w$. Hence, if $M$ is a Turing-machine deciding
$P$, we can now use the formula $\phi_M$ constructed in
Lemma~\ref{lem:TM-enc} such that $H\models \phi_M$ if, and only, if
$M$ accepts the word $w$ if, and only if, $w\in P$. By definition of
transductions, this gives us a formula $\psi_M$ which is true in $G_w$
if, and only if, $\phi_M$ is true in $H$ if, and only if, $w\in P$.

Now, using the conditions $1$ and $2$ of $(f, p)$-unboundedness, we
get that we can always find such graphs $G$ and $G_w$
efficiently. Furthermore, as $\CCC$ is closed under
colourings, $G_w$ is also in $\CCC$. Hence, if $\pMC(\MSO, \CCC)$ was
in XP, i.e.~$G_w\models \psi_M$ could be decided in time
$|G_w|^{f(|\psi_M|)}$, then $P$ could be decided in polynomial time as
$\phi_M$ does not depend on the input $w$ and the exponent is
therefore fixed.

The problem with this approach is that the tree-width of $G$ is
only logarithmic in $|G|$ and hence $G$, and thus $G_w$, can be of size
exponential in its tree-width. Furthermore, the best known bound for
the size of grids we are
guaranteed to find by the excluded grid theorem is only logarithmic
in the tree-width of the graph. Hence, in order to guarantee that $G$
contains a grid of size $|w|$ we would need to construct a graph of
tree-width exponential in the length $|w|$ of $w$ which could therefore be of double
exponential size in $|w|$. This completely destroys the argument above,
as deciding $G_w\models \psi_M$ in time
$|G_w|^{f(|\psi_M|)}$ only yields that we can decide $w\in P$ in time
doubly exponential in $w$ and  this certainly can be done for
$\NP$-problems. 

To get the result we want, we need to find grids of size
polynomial in the tree-width of $G$. For, suppose for every graph $G$
we could find a grid of size polynomial in its tree-width. Then, given
$w$ we could use the conditions of $(f, p)$-unboundedness to construct a
graph $G$ of tree-width polynomial in $w$, and hence containing a grid
of size $|w|\times|w|$, whose size is bounded by $2^{o(|w|)}$ (this
will be explained in detail in Section~\ref{sec:main}). We could then
colour this grid to encode $w$ as before to obtain $G_w$. Now, if 
$G_w\models\psi_M$ could be decided in time $|G_w|^{f(|\psi_M|)}$,
then this would imply that $w\in P$ could be decided in time
$\big(2^{o(w)}\big)^{f(|\psi_M|)}$ which is the same as
$\big(2^{{f(|\psi_M|)}\cdot o(w)}\big)$ and hence in time
sub-exponential in $w$. And sub-exponential solvability of
$\NP$-complete problems in case $\pMC(\MSO, \CCC) \in $ XP is exactly
what we claim in Theorem~\ref{theo:main}.

Obtaining sub-exponential time algorithms for problems such as TSP or
\textsc{Sat} is an important open problem in complexity theory
and the common assumption is that no such algorithms exist. 
This has led to the \emph{exponential-time hypothesis}
(ETH) which says that there is no such sub-exponential time algorithm
for \textsc{Sat}, a hypothesis widely believed in the community.

Hence, to prove our main result we need to find grids of size
polynomial in the tree-width of graphs. The existence of such grids is
a major open problem in structural graph theory and remains open to
date.
Instead of grids we will therefore use a replacement structure for
grids, called \emph{grid-like minors}, recently introduced by Reed and
Wood \cite{ReedWoo08}. A grid-like minor of order $l$ in a graph $G$ is a pair
$\PPP, \QQQ$ of sets of pairwise disjoint paths such that their
intersection graph $\III(\PPP, \QQQ)$ contains an $l\times l$-grid as
a minor. It was shown in \cite{ReedWoo08} that every graph $G$
contains a grid-like minor of order polynomial in its tree-width (see
the next section for details). 

Our method for proving Theorem~\ref{theo:main} is therefore exactly as
outlined above, only that instead of defining grid-minors and
colouring them appropriately, we will define grid-like minors and
colour those appropriately. This, however, is significantly more
complicated than the  case of grid-minors.

One of the problems is that the grid-like minor is actually a
grid-minor of the intersection graph of two sets of pairwise disjoint
paths. Hence, to define it in $\MSO$ we will have to define these sets
of disjoint paths, then define their intersection graph and then
define a grid-minor in it. This already is somewhat more complicated
than defining pure grids.

The second, and major, challenge is to colour this grid-like minor so
that it encodes a word $w$. For this, we need to colour the vertices
and edges of the graph $G$ so that this induces an appropriate
colouring of the grid-minor of the intersection graph of two sets of
disjoint paths. All this needs to be done in a way that once we have
coloured the vertices and edges of $G$, there are no two different
grid-like minors in $G$ for which the colouring induces different words. 

For this, we will define a combinatorial structure, called
pseudo-walls, and show that every graph $G$ contains a pseudo-wall of
order polynomial in the tree-width of $G$, that we can colour the graph
$G$ in a way that it induces a unique colouring of this pseudo-wall,
that we can define the pseudo-wall in $\MSO$ 
and, finally, that we can define an appropriately coloured grid in
this pseudo-wall. Pseudo-walls and their colourings are defined in
Section~\ref{sec:pseudo-walls}.
Definability of these structures in $\MSO$ is proved in
Section~\ref{sec:msodef}. Finally, we complete the proof in Section~\ref{sec:main}.

\section{Pseudo-Walls in Graphs}
\label{sec:pseudo-walls} 

This section contains the graph theoretical and algorithmic aspects of
the proof outlined in the previous section. We first define the
notions of simple and complex pseudo-walls and show that any graph of
large enough tree-width can be expanded to a $\sigmacol$-structure
containing either a simple or complex pseudo-wall of large order. 

\begin{figure}
  \centering
  \includegraphics[height=3cm]{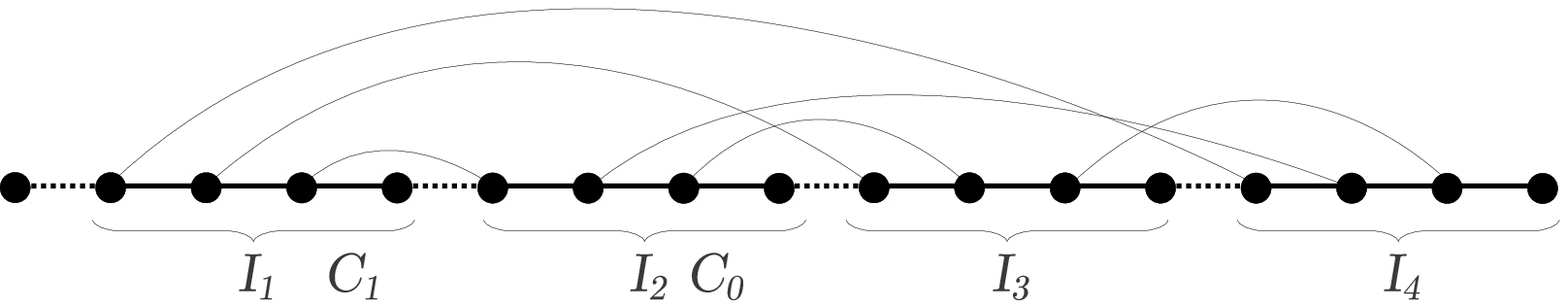}
  \caption{Simple pseudo-wall}
  \label{fig:simple-pw}
\end{figure}

A simple pseudo-wall is a structure as illustrated in
Figure~\ref{fig:simple-pw}. Essentially, it consists of a long path $L$ whose edges are
coloured either red (solid horizontal lines in the figure) or blue
(dashed lines) together with a set $\QQQ$ of pairwise
vertex-disjoint paths (represented by the curved lines in the figure).
 The first edge of $L$ is blue but the
last is red so that this gives the path a direction. Furthermore, the
blue edges partition the path into segments formed by the red
edges and for any pair of such segments there is a path in $\QQQ$
linking them. All vertices in a segment have
the same colour with respect to $C_0, C_1$, i.e.~they are either all
in $C_1$ or all in $C_0$ or all uncoloured. Finally, the vertices
coloured by $C_0 \cup C_1$ occur to the left of the long path
$L$. This will allow us to define a coloured clique from a simple
pseudo-wall where the vertices of the clique are formed by the red
segments of $L$ and the edges are defined by the paths in $\QQQ$.
Formally, a simple pseudo-wall is defined as follows.

\begin{defi}[Simple Pseudo-Wall]\label{def:simple-pw}
  A \emph{simple pseudo-wall of order $k$} is a $\sigmacol$-structure
  $\AA := (A, 
  V^\AA, E^\AA, \in^\AA, B^\AA, R^\AA, C_0^\AA, C_1^\AA)$ defined as follows.
  Let $h := ({k \atop 2})$.

  \begin{iteMize}{$\bullet$}
  \item $A := \{ v_1, \dots, v_{s+1}, e_1, \dots, e_s, \} \cup 
    \{ u^i_j, e^i_l \st 1\leq i \leq h, 1\leq j\leq s_i+1, 1\leq
    l\leq s_i\}$ for some $s, s_1, \dots, s_h >0$.
  \item $V^\AA := \{ v_1, \dots, v_{s+1}, u^i_j \st 1\leq i\leq h,
    1\leq j\leq s_i+1\}$.
  \item $E^\AA := \{ e_1, \dots, e_{s}, e^i_j \st 1\leq i\leq h,
    1\leq j\leq s_i\}$.
  \item $\in^\AA := \{ (a,b) \st a\in V, b\in E, a\in b\}$.
  \item 
    $L := (v_1, e_1, v_2, \dots, e_s, v_{s+1})$ forms a path of length $s$.
  \item There is a tuple $I := (i_1, \dots, i_{k})$ of indices
    $1\leq i_j\leq s$, for $1\leq j\leq k$, such that $i_1 :=
    1$, $i_j < i_{j+1} < i_j+h$, for all $1\leq j < h$, and $i_{h} <
    s$.  For all $1\leq j\leq k$, we call $\{ v_{i_j+1}, \dots,
    v_{i_{j+1}} \}\subseteq V(P)$ the \emph{$j$-th interval $\III_j$
      of $P$}, where  we set $i_{h+1} := s+1$. Then $R^\AA := \{ e_l
    \st l \not\in I\}$ and $B^\AA := \{ e_{i_j} \st i_j\in I\}$.
  \item $C_0^\AA, C_1^\AA \subseteq \{ v_1, \dots,
    v_{s+1}\}$ are pairwise disjoint sets such that for all $1\leq
    j\leq k$ and $C \in \{ C_0, C_1 \}$, either $\III_j \subseteq
    C^\AA$ or $\III_j \cap C^\AA = \emptyset$. Furthermore, for all
    $j$, if $\III_j \subseteq C_0 \cup C_1$ and $i<j$ then $\III_i
    \subseteq C_0 \cup C_1$. 
  \item $\QQQ := \{ (u^i_1, e^i_1, \dots, e^i_{s_i}, v^i_{s_i}) \st 1
    \leq i \leq h\}$ forms a set of pairwise disjoint paths
    $P_{i,j}$, $1\leq i < j\leq k$, 
    such that $P_{i,j}$ links $\III_i$ and $\III_j$, i.e.~$u^i_1 \in
    \III_i$, $u^i_{s_i+1}\in \III_j$ and $u^i_j \in^\AA e^i_j$ and
    $u^i_j\in^\AA e^i_{j-1}$ for all suitable $j$.
  \end{iteMize}
  Let $l\leq k$ be maximal with $\III_l\subseteq C_0 \cup C_1$. 
  The \emph{word $w$ encoded by $\AA$} is the sequence $w := w_1,
  \dots, w_l \in \{ 0,1\}^*$ with $w_i := 1$ if  $\III_i
  \subseteq C^\AA_1$ and  $w_i := 0$ if $\III_i \subseteq C^\AA_0$.
\end{defi}

Note that the intersection graph of the set $\QQQ$ and the set of
paths comprising the intervals forms a complete graph on $k$
vertices. The
colouring of intervals by $C_1$ and $C_0$, respectively, yields a
colouring of this clique in an obvious way. We will show in the next
section that  if a
$\sigmacol$-structure $\BB$ contains such a simple pseudo-wall $\AA$
encoding a word $w$ as sub-structure, we can use this in a similar
way to Section~\ref{sec:mso} to simulate the run of a Turing machine
on input $w$. 

However, we may not always be able to find sufficiently large simple
pseudo-walls in a $\sigmacol$-structure. Instead we may have to settle
for a more complicated structure, called complex pseudo-walls.

\begin{defi}[Complex Pseudo-Wall]
  A \emph{complex pseudo-wall of order $k$} is a $\sigmacol$-structure
  $\AA := (A, V^\AA, E^\AA, \in^\AA, B^\AA, R^\AA, C_0^\AA, C_1^\AA)$
  defined as follows.
  
  \begin{iteMize}{$\bullet$}
  \item $A := \{ v_1, \dots, v_{s+1}, e_1, \dots, e_s, \} \cup 
    \{ u^i_j, e^i_l \st 1\leq i \leq r_1+r_2, 1\leq j\leq s_i+1, 1\leq
    l\leq s_i\}$ for some $r_1, r_2, s, s_1, \dots, s_r >0$.
  \item $V^\AA := \{ v_1, \dots, v_{s+1} \} \cup \{ u^i_j \st 1\leq
    i\leq r_1+r_2, 1\leq j\leq s_i+1\}$.
  \item $E^\AA := \{ e_1, \dots, e_s \} \cup \{ e^i_j \st 1\leq
    i\leq r_1+r_2, 1\leq j\leq s_i\}$.
  \item $L := (v_1, e_1, v_2, \dots, e_s, v_{s+1})$ forms a path of length
    $s$.
  \item $B^\AA := \{ e_1 \}$.
  \item $R^\AA := \{ e_2, \dots, e_s \}$.
  \item $C_0^\AA, C_1^\AA \subseteq \{ v_1, \dots, v_{s+1}\}$ and
    $C_0^\AA \cap C_1^\AA = \emptyset$.
  \item  $\PPP := \{ (u^i_1, e^i_1, \dots, e^i_{s_i}, v^i_{s_i}) \st 1
    \leq i \leq r_1\}$ and $\QQQ := \{ (u^i_1, e^i_1, \dots,
    e^i_{s_i}, v^i_{s_i}) \st r_1 
    < i \leq r_2\}$  form sets of pairwise disjoint paths
    $P_i := (v^i_1, e^i_1, \dots, u^i_{s_i+1})$ and $Q_i := (u^i_1, e^i_1, \dots,
    e^i_{s_i}, v^i_{s_i})$ so that every path
    $P\in\PPP$ intersects $L$ in one endpoint of $P$ but has no other
    vertex with $L$ in common.

    Furthermore, $\III(\PPP,
    \QQQ)$ contains an image $\mu$ of a complete graph $K_{k^2}$ as 
    topological minor such that if $U := \{ v_1, \dots, v_{s+1}\} \cap
    (C_0^\AA \cup C_1^\AA)$ then for each $u\in U$ there is a branch
    set $\mu_u$ containing a path $P\in \PPP$ with one endpoint
    being $u$ and if $u\not=u'\in U$ then $\mu_u \cap \mu_{u'} = \emptyset$.
  \end{iteMize}
  Let $i_1, \dots, i_n$ be the indices of the vertices $v_i \in
  C_0^\AA\cup C_1^\AA$. 
  The \emph{word $w$ encoded by $\AA$} is $w := w_1, \dots, w_n$ where
  $w_j := 1$ if $v_{i_j} \in C_1^\AA$ and $w_j := 0$ if $v_{i_j} \in
  C_0^\AA$. 
\end{defi}

\begin{figure}
  \centering
  \includegraphics[height=4cm]{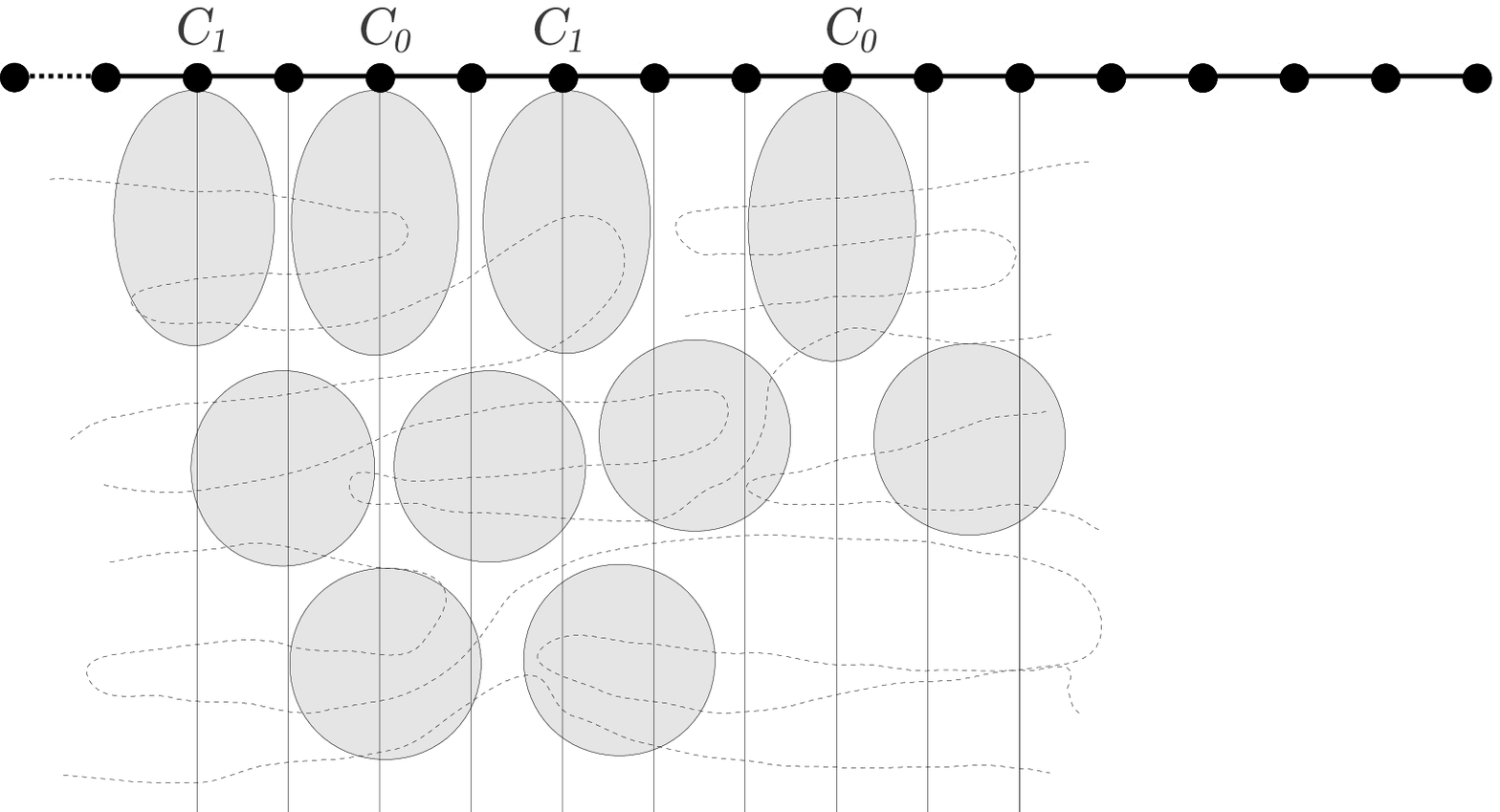}
  \caption{Complex pseudo-wall}
  \label{fig:complex-pw}
\end{figure}

Figure~\ref{fig:complex-pw} illustrates a complex pseudo-wall encoding
the word $1010$. Here, the horizontal lines and bullets form the path
$P$. The dashed line at the top-left indicates the ``blue'' edge $e_1$
and the horizontal solid lines the other edges $e_2, \dots, e_s$. The
vertical lines indicate the paths in $\PPP$ and the curved lines the
paths in $\QQQ$. The
grey areas represent the branch sets of the clique minor. Note that
the figure is only an illustration as the paths in $\QQQ$ as displayed
in the figure do not generate an intersection graph with a large
clique-minor as required by complex pseudo-walls. 

The motivation behind complex pseuo-walls is that the path $P$ is used
to encode a word $w$. 
The ``blue'' edge $e_1$ only serves the purpose of giving the path $P$
an orientation, with $e_1$ marking the left end of $P$ so that the
word encoded in the wall is always read in the correct order. The sets
$\PPP$ and $\QQQ$ form an intersection graph containing a topological
clique minor. The
requirement that every coloured vertex occurs in exactly one
branch set of this minor ensures that we can assign colours to the
branch sets and therefore,
given a complex pseudo-wall, we
can define from it a vertex coloured clique. Furthermore, we can
define an order on the vertices of this clique induced by the order
defined by the path $L$. This will be enough to define a coloured grid
in this clique which encodes the same word as the original complex
pseudo-wall. Details of this construction will be given in the next section.

\begin{defi}[pseudo-walls] A $\sigmacol$-structure $\AA$ is a
  \emph{pseudo-wall of order $k$ encoding a word $w$} if it a simple
  or complex pseudo-wall of order $k$ encoding $w$.  
\end{defi}

We will see later that pseudo-walls in
$\sigmacol$-structures can be defined in $\MSO$.
The main result of this section is the following theorem showing that any graph $G$ can be
expanded, or coloured, to a $\sigmacol$-structure $\AA$ containing a
pseudo-wall of order polynomial in the tree-width of $G$.

\begin{thm}\label{theo:pseudo-wall}
  There is a polynomial-time algorithm and a constant $c$ such that
  given a graph $G$ such that 
  \[
    \big(\frac{\tw(G)}{\sqrt{\log \tw(G)}}\big)^{\frac13} \geq c\cdot m^7+1
  \]
  and a word
  $w\in \{ 0,1\}^*$ of length at most $m$ computes a
  $\sigmacol$-expansion of $G$ containing a pseudo-wall of order $m$
  encoding $w$.
\end{thm}

In \cite{ReedWoo08} Reed and Wood consider an alternative to
grid-minors as obstructions to small tree-width which they call
\emph{grid-like} minors. A 
grid-like minor of order 
$l$ in a graph $G$ is a set $\PPP$ of paths in $G$ such that the
intersection graph 
$\III(\PPP)$ contains a $K_l$-minor. 
Reed and Wood's proof is existential, in that
it does not directly give a way of computing grid-like minors. In
\cite{KreutzerTaz10}, Kreutzer and Tazari show that the
individual parts of this 
proof can be made algorithmic and a polynomial-time algorithm for
computing grid-like minors is given. 

Grid-like minors are the key to finding pseudo-walls. However, we
cannot use Reed and Wood's result directly but have to adapt their
proof slightly to get the structures we need. The following is
essentially the proof from \cite{ReedWoo08} and the algorithmic
components from \cite{KreutzerTaz10} needed to make it algorithmic,
suitably adapted to yield pseudo-walls instead of grid-like minors.

The starting point of the proof are brambles. By
Theorem~\ref{theo:bramble}, every graph $G$ contains a bramble of
order $\tw(G)+1$. However, these can be of size exponential in $|G|$ and, as
proved by Grohe and Marx \cite{GroheM09}, there is an infinite family
of graphs where 
brambles of optimal order necessarily are of exponential
size. However, if we settle for brambles whose order is only
polynomial in the tree-width, polynomial size can always be
guaranteed. The existence of such brambles was proved in
\cite{GroheM09}, a polynomial-time algorithm for computing them was
given in \cite{KreutzerTaz10}.

\begin{thm}[\cite{GroheM09,KreutzerTaz10}]\label{theo:bramble-alg}
  There exists a polynomial time algorithm which, given a graph $G$, 
  constructs a bramble in $G$ of size
  $O(\tw(G))$ and order $\Omega((\frac{\tw(G)}{\sqrt{\log
  \tw(G)}})^{1/3})$.
\end{thm}

We first need the following lemma, whose simple proof is included 
for the reader's convenience.

\begin{lem}[\thmcite{Birmel\'e, Bondy, Reed}{\cite{BirmeleBonRee07}}]
  \label{lem:longpath}
  Let $\BBB$ be a bramble in a graph $G$. Then $G$ contains a path
  intersecting every element in $\BBB$.
\end{lem}
\begin{Proof}
  Choose a bramble element $B\in\BBB$ and a vertex $v\in
  \vrtB$. We initialise a path $P := (v)$ and maintain the invariant that for one
  endpoint $u$ of $P$ there is a bramble element $B\in\BBB$ such that
  $\vrtB \cap \vrtP = \{ u \}$. The invariant trivially holds for $P
  = (v)$. So suppose such a path $P$ has been constructed and let $u$
  be the endpoint of $P$ as stated in the invariant. While there still is a bramble element
  $B' \in\BBB$ not containing a vertex of $P$ choose a path $P'$ from
  $u$ to $B'$ in $B\cup B'$ as short as possible. Such as path exists as
  $B$ and $B'$ touch. As $P'$ is chosen as short as possible, one
  endpoint of $P'$ is the only element of $P'$ in $B'$. Further, as $u$
  is the only element of $P$ in $B$, $P\cdot P'$, i.e.~the path obtained
  from adding $P'$ to $P$ at the vertex $u$ is still a path satisfying
  the invariant. We proceed until there are no bramble elements left
  which have an empty intersection with $P$.
\end{Proof}

Clearly, if $P$ is a path in $G$ intersecting every element of a bramble $\BBB$
then the length of $P$ must be at least the order of $\BBB$.

\begin{lem}[\thmcite{Reed and Wood}{\cite{ReedWoo08}}]\label{lem:paths}
  Let $G$ be a graph containing a bramble $\BBB$ of order at least $kl$, for
  some $k,l\geq 1$. Then $G$ contains $l$ pairwise vertex disjoint
  disjoint paths $P_1,\dots, P_l$ s.t.  
  for all $1 \leq i < j \leq l$, $G$ contains $k$ parwise vertex disjoint
  paths between $P_i$ and $P_j$.
\end{lem}
\begin{Proof}
  By Lemma~\ref{lem:longpath}, there is a path $P := (v_1 \dots v_n)$ in $G$
  intersecting every element of $\BBB$ and hence of length at least $kl$.
  For $1\leq i\leq j\leq n$ let $P_{i,j}$ be the sub-path of $P$ induced by
  $\{v_i, \dots, v_j\}$. Let $t_1$ be the minimal integer such that the
  sub-bramble $\BBB_1 := \{ B\in\BBB \st B\cap P_{1,t_1}\not=\emptyset\}$ has
  order $k$. Given $t_i, \BBB_i$ with $i<l$, let $t_{i+1}$ be the minimal
  integer such that the sub-bramble 
  $\BBB_{i+1} := \{ B\in\BBB \st B\cap P_{t_i+1,t_{i+1}}\not=\emptyset, B\cap
  P_{1,t_i} = \emptyset\}$ has order $k$. Since $\BBB$ has order $kl$,
  in this way we obtain
  integers $t_1 < t_2 < \dots < t_l \leq n$. Let $P_i :=
  P_{t_{i-1}+1,t_i}$, where $t_0 := 0$. By construction, the $P_i$ are pairwise
  disjoint.
  
  Suppose there is a set $S\subseteq \vrtG$ of cardinality $|S|<k$ separating
  some $P_i$ and $P_j$. Hence, $S$ is neither a hitting set of $\BBB_i$ nor of
  $\BBB_j$ and hence there is $B_i\in\BBB_i$ and $B_j\in\BBB_j$ such that
  $S\cap (B_i\cup B_j)=\emptyset$. As $B_i$ and $B_j$ touch it follows that $S$
  does not separate $B_i$ and $B_j$ and therefore does not separate $P_i$ and
  $P_j$. Hence, any set separating $P_i$ and $P_j$ must be of cardinality at
  least $k$. 

  By Menger's theorem~\ref{thm:menger}, the minimal cardinality of a
  set separating $P_i$ and $P_j$ is equal to the maximum number of
  pairwise vertex disjoint paths between $P_i$ and $P_j$, and hence
  there are at least $k$ pairwise vertex disjoint paths between $P_i$
  and $P_j$ as required. 
\end{Proof}

A graph $G$ is \emph{$d$-degenerated} if every
subgraph of $G$ contains a vertex of degree at most $d$. Mader \cite{Mader67}
proved that every graph with no $K_l$-minor is $2^{l-2}$-degenerated. Let $d(l)$
be the minimal integer such that every graph with no $K_l$-minor is
$d(l)$-degenerated. Kostocha and, independently,
Thomason showed that $d(l) \in \oldtheta(l\sqrt{\log l})$.  
Bollob{\'a}s and Thomason \cite{BollobasTho98} proved that there is a constant $c$ so that if a
graph has average degree at least $c p^2$ it contains a $K_p$ as a
topological minor. Here we need an algorithmic version of this result,
proved in \cite{KreutzerTaz10}. 

\begin{thm}[\cite{BollobasTho98,KreutzerTaz10}]\label{theo:top-minor}
  There is a constant $d$ such that if a graph $G$ has average degree
  at least $dp^2$, then $G$ contains $K_p$ as a topological minor.
  Furthermore, a model of $K_p$ in $G$ can be found in polynomial time.
\end{thm}

We are now ready to prove Theorem~\ref{theo:pseudo-wall}. 

\medskip

\begin{Proof}[Proof of Theorem~\ref{theo:pseudo-wall}]
  Set $c := d$ where $d$ is the constant from Theorem~\ref{theo:top-minor}.
  Let $k := ({m \atop 2})\cdot\big(({m \atop 2})-1\big)\cdot d\cdot m^2+1$.
  Let $w := \big(\frac{\tw(G)}{\sqrt{\log
      \tw(G)}}\big)^{\frac13}$. Then $w\geq k\cdot m+1$.

  By
  Theorem~\ref{theo:bramble-alg}, we can compute in polynomial time a
  bramble $\BBB$ in 
  $G$ of order at least
  $k\cdot m+1$. Therefore, by Lemma~\ref{lem:paths}, $G$ contains a path $A$
  of length $k\cdot m+1$. Fix one endpoint $p$ of $A$ and let $e_0$ be the
  unique edge of $A$ incident to $p$. Then, $A\setminus
  \{p\}$ has length at least $k\cdot m$ and can be decomposed into $m$
  disjoint paths $P_1, \dots, P_m$ and, 
  for $1 \leq i < j \leq m$, $G$ contains a set $\QQQ_{i,j}$ of $k$
  disjoint paths between $P_i$ and $P_j$. The edge $e_0$ needs to be
  set aside for the case of simple pseudo-walls below.
  
  For $1 \leq i < j \leq m$ and $1 \leq a < b \leq m$ such that
  $\{i,j\}\not=\{a,b\}$, let $H_{i,j,a,b} := \III(\QQQ_{i,j},
  \QQQ_{a,b})$ be the intersection graph of $\QQQ_{i,j} \cup
  \QQQ_{a,b}$.
 
  \smallskip

   \noindent\textit{Complex Pseduo-Walls. }
  Suppose there are $i,j,a,b$ as above such that $H_{i,j,a,b}$ has
  a sub-graph of average degree at least $d\cdot m^2$. We define a
  $\sigmacol$-expansion $\AA$ of $G$ which contains a complex
  pseudo-wall of order $m$ encoding $w$ as follows.

  By Theorem~\ref{theo:top-minor}, $H := H_{i,j,a,b}$ contains a $K_m$ as
  topological minor and we can compute an image of it in polynomial
  time. Set $L := P_i$. Fix one endpoint of $L$ and let $e_1$ be the
  edge incident to it in $L$. We define $B^{\AA}
  := \{ e_1 \}$ and $R^{\AA} := E(L) \setminus \{e_1\}$. This defines
  a direction on $L$ where the endpoint incident to $e_1$ is the
  left-most, or smallest. 

  This direction induces an ordering $\sqsubset$ on the
  paths in $\QQQ_{i,j}$ where for $P, P'\in \QQQ_{i,j}$ we define $P\sqsubset P'$ if $V(P) \cap V(L)$ is
  smaller than $V(P')\cap V(L)$. (Note that any $P\in \QQQ_{i,j}$ has
  exactly one vertex in common with $L$, which is its endpoint in $L$.)

  Let $X_1, \dots, X_m$ be the connected subgraphs
  in $H$ constituting the image of $K_m$ in
  $H$. W.l.o.g.~we assume that each $X_i$ contains a path from
  $\QQQ_{i,j}$. (There can only be at most one $X_i$ consisting of a single
  path from $\QQQ_{a,b}$.) 
  For each $1\leq i \leq m$ let $p_i$ be the smallest vertex in $V(L)$
  with respect to $\sqsubset$ contained in a path in $X_i$. 
  We order the sets $X_i$ by letting $X_i < X_j$ if $p_i \sqsubset p_j$.
  W.l.o.g.~we assume that $X_1 < X_2 < \dots < X_m$.
  Then, $C_0^\AA := \{ p_i \st 1 \leq i \leq |w|, w_i = 0\}$ and
  $C_0^\AA := \{ p_i \st 1 \leq i \leq |w|, w_i = 1\}$, where $w :=
  w_1, \dots, w_{|w|}$

  It is now immediately clear from the construction, that $\AA$ contains a
  complex pseudo-wall of order $m$ encoding $w$: the wall is
  constituted by $L$, $\PPP := \QQQ_{i,j}$ and $\QQQ :=
  \QQQ_{a,b}$ and the colours $R^\AA, B^\AA, C_0^\AA, C_1^\BB$.

  \smallskip

   \noindent\textit{Simple Pseduo-Walls. }
   Now suppose that the average degree of all sub-graphs of
   $H_{i,j,a,b}$, where  $\{i,j\} \not= \{a,b\}$, is less than $d\cdot
   m^2$, i.e.~all $H_{i,j,a,b}$ are $d\cdot m^2$-degenerated. 

   Let $H$ be the intersection graph of $\bigcup \{
  \QQQ_{i,j} \st 1\leq i < j \leq m\}$. Obviously, $H$ is ${m\choose
    2}$-colourable with each $\QQQ_{i,j}$ being a colour class. Each
  colour class has $k$ vertices and each pair of colour classes induce a
  $d\cdot m^2$-degenerated graph. The following lemma is from \cite{ReedWoo08}.
  
  \begin{lem}[\cite{ReedWoo08}]
      Let $r\geq 2$ and let $V_1, \dots, V_r$ be the colour classes in
      an $r$-colouring of a graph $H$. Suppose that $|V_i|\geq n :=
      r(r-1)c + 1$, for all $1 \leq i \leq r$, and $H[V_i \cup V_j]$
      is $c$-degenerated for distinct $1 \leq i < j \leq r$. Then
      there exists an independent set $\{ x_1, \dots, x_r\}$ of $H$
      such that each $x_i \in V_i$. 

      Furthermore, a simple
      minimum-degree greedy algorithm will find such an independent
      set in polynomial time. 
  \end{lem}
  
  Applying the lemma to our setting, with $n=k$ and $r= {m\choose 2}$
  and $c = d\cdot m^2$, we obtain an independent set $I$ in $H$ with one vertex in
  each colour class and such a set can be found in polynomial time by a
  simple greedy algorithm. That is, in each set $\QQQ_{i,j}$ there is one path
  $Q_{i,j}$ such that $Q_{i,j} \cap Q_{a,b} =\emptyset$ whenever
  $\{i,j\} \not= \{a,b\}$.

  We will now define a $\sigmacol$-expansion $\AA$ of $G$ containing a
  simple pseudo-wall of order $m$ encoding $w$. 
  Consider the long path $A$ constructed above. Recall
  that there is one edge $e_0$ of $A$ incident to an endpoint $p$ and
  that $L\setminus \{p\}$ is partitioned into $P_1, \dots, P_m$. As
  $P_1, \dots, P_m$ are pairwise disjoint, between any $P_i$ and
  $P_{i+1}$ there is one edge $e_i$ of $L$ not contained in $P_i\cup
  P_{i+1}$. Let $B^\AA := \{ e_0, e_1, \dots e_{m-1}\}$ and $R^\AA :=
  E(A)\setminus B^\AA$. Furthermore, if $w := w_1, \dots, w_l$, with
  $l\leq m$, then $C_0 := \bigcup \{ V(P_i) : w_i = 0\}$ and $C_1 :=
  \bigcup \{ V(P_i) : w_i = 1\}$. 

  By construction, $\AA$ contains a simple pseudo-wall of order $m$
  encoding $w$, which is generated by the long path $A$, the colours
  $B^\AA, R^\AA, C_0^\AA, C_1^\AA$ and the paths in $I$. 

  This concludes the proof of Theorem~\ref{theo:pseudo-wall}.
\end{Proof}

\section{Intractability of $\MSO$ on Pseudo-Walls}
\label{sec:msodef}

The main purpose of this section is to show that $\MSO$ is intractable
on the class of pseudo-walls. For this purpose, we will lift
Lemma~\ref{lem:TM-enc} from grids to pseudo-walls. 

To get the result we will exhibit a sequence of $\MSO$-transductions
that define coloured grids in pseudo-walls. 
To simplify the
presentation, we will do so in several steps. Obviously,
the transductions will be different for simple and complex
pseudo-walls. The sequence of transductions works as follows. We will
first exhibit a transduction defining coloured grids in coloured
ordered cliques. We will then show that there are transductions
defining coloured ordered cliques in simple and complex pseudo-walls,
where in the latter we will need one further intermediate step.

\subsection{Coloured ordered cliques. }
  Recall from Definition~\ref{def:signatures} the signatures
  $\sigmaord$ and $\sigmaw$.
\begin{defi}
  
  A \emph{coloured ordered clique} is a $\sigmaord$-structure $\AA :=
  (U, V, E, \in, C_0, C_1, \leq)$ so that
  \begin{iteMize}{$\bullet$}
  \item $(U,
    V, E, \in)$ is the incidence representation of a complete graph
  \item $\leq$ is a linear order on $V$ and
  \item $C_0, C_1 \subseteq V$, $C_0 \cap C_1 = \emptyset$ and $C_0 \cup C_1$ forms an initial
    subset of $\leq$, i.e.~there is a $v\in C_0 \cup C_1$ such that
    $C_0 \cup C_1 = \{ u\in V \st u \leq v\}$.
  \end{iteMize}
  The \emph{order} of $\AA$ is $|V|$.
  Let $v_1, \dots, v_n$ be the vertices in $C_0 \cup C_1$ ordered by
  $\leq$. 
  The word $w$ \emph{encoded by $\AA$} is $w :=   w_1, \dots, w_n$
  where $w_i := 1$ if $v_i\in C_1$ and $w_i := 0$ if $w_i \in C_0$. 
\end{defi}

\begin{lem}\label{lem:clique-trans}
  There is an $\MSO$-transduction $\Theta$ from $\sigmaord$ to $\sigmaw$
  with parameters $\PPP, \QQQ$ such that if $\AA$ is a coloured
  ordered clique of order $k$ encoding a word $w$ then $\Theta(\AA)$
  contains a coloured $(\sqrt{k}\times \sqrt{k})$-grid encoding
  $w$. Furthermore, every $\BBB\in \Theta(\AA)$ is a grid encoding
  $w$. 
\end{lem}
\begin{Proof}
  We define the transduction $\Theta := (\phivalid,
  \phiuniv, \phi_\sim, \phi_V, \phi_E, \phi_\in, \phi_{C_0},
  \phi_{C_1})$ as follows. 

  The transduction is quite simple as the grid we seek to define is actually a
  sub-structure of the given coloured clique. The idea is that the
  parameters $\PPP, \QQQ$ will be enforced to be interpreted by two
  sets of pairwise vertex disjoint paths, the \emph{vertical} and
  \emph{horizontal} paths in a grid. All we need to say is that they
  indeed form a grid, that the bottom row of the grid contains all
  coloured vertices from left to right in the order given by $\leq$.

  So let $\phi_V(x) := x \in V$, $\phi_E(x) := x \in E$ and $\phiuniv(x) :=
  \phi_V \vee \phi_E$. We set $\phi_\sim(x, y) := x=y$.  Furthermore,
  we define $\phi_{C_0}(x) := x \in C_0$ and $\phi_{C_1}(x) := x \in
  C_1$. 
  What
  is left to define is $\phivalid$. Recall the formula
  $\phi_{\textit{grid-border}}(L, R, T, B, \PPP, \QQQ)$ from
  Example~\ref{ex:grid} defining that $\PPP, \QQQ$ are two sets
  of pairwise vertex disjoint paths inducing a grid such that bottom,
  left, top, and right rows are $B, L, T, R$, respectively. 
  We will also use the formulas $\maxpath, \phiendpoint$ and $\sodp$ defined in this
  example.

   $\phivalid$ will enforce $\PPP, \QQQ$ to be interpreted by sets of
  pairwise disjoint paths inducing a grid whose bottom row contains
  the coloured vertices in the correct order. 
  As before we will therefore use the
  notation $Q\in \QQQ$ as shorthand for $Q\subseteq \QQQ \wedge
  \maxpath(Q,\QQQ)$.

\begin{eqnarray*}
  \phivalid(\PPP, \QQQ) &:=& \sodp(\PPP) \wedge \sodp(\QQQ) \ \wedge
  \\
  &&\exists B, T \in \QQQ\ \exists L, R \in \PPP\ \big[\phi_{\textit{grid-border}}(\PPP,
  \QQQ, B, T, R, L) \ \wedge\\
  && \exists x \in V(B) x \in C_0 \cup C_1 \wedge \forall y (y \in
  C_0 \cup C_1 \leftrightarrow y \in V(B) \wedge y \leq x)\ \wedge\\
  &&  \forall x, y \in V(B)\big( \phi_{\leq_B}(x, y) \leftrightarrow x \leq y\big)\big]
\end{eqnarray*}
where 
\begin{eqnarray*}
  \phi_{\leq_B}(x, y, B)& := &\exists u \in V(B) \wedge
  \phiendpoint(u, B) \wedge \forall y u\leq y\ \wedge\ \\&& 
  \forall P\subseteq B(path(P, B) \wedge u\in V(P) \wedge y\in
  V(P) \rightarrow x\in V(P)).
\end{eqnarray*}

The formula $\phi_{\leq_B}$ states that one endpoint $u$ of $B$ is the
$\leq$-smallest element in the structure and then defines a linear
order on $B$ where $x$ is smaller than $y$ if the distance from
$x$ to $u$ in $B$ is smaller than the distance from $y$ to $u$. This is
formalised by stating that any sub-path of $B$ which contains $u$ and
$y$ must also contain $x$.

$\phivalid$ then states that $\PPP, \QQQ$ are sets of pairwise
disjoint paths defining a grid with bottom row $B\in \QQQ$ and that
the vertices in $C_0$ and $C_1$ all occur as an initial subpath on $B$
in the order given by $\leq$. Hence, this grid encodes the same word
as the initial structure.

This shows that every structure in $\Theta(\AA)$ is a grid encoding
$w$. Furthermore, if $\PPP, \QQQ$ are chosen as the vertical and
horizontal paths in a $\sqrt{k}\times \sqrt{k}$-grid which exists as a
sub-structure of the clique $\AA$, then $\Theta((\AA, \PPP, \QQQ))$
has order $\sqrt{k}\times\sqrt{k}$. This concludes the proof. 
\end{Proof}
\begin{cor}\label{cor:TM-clique}
  Let $M$ be a non-deterministic $n^d$-time bounded Turing-machine.
  There is a formula $\phi_M\in\MSO$ such that for all words
  $w\in\Sigma^\star$, if $G$ is a coloured ordered clique of order
  $|w|^d$  encoding
  $w$, then $G\models \phi_M$ if, and only if, $M$
  accepts $w$.  Furthermore, the formula $\phi_M$ can be constructed
  effectively from $M$.  

  The same holds if $M$ is an alternating
  Turing-machine with a bounded number of alternations, as they are
  used to define the polynomial-time hierarchy.
\end{cor}
\begin{Proof}
  The corollary follows immediately from Lemma~\ref{lem:TM-enc},
  Corollary~\ref{cor:inter} and the previous
  Lemma~\ref{lem:clique-trans}. 
\end{Proof}

\subsection{Simple Pseudo-Walls}

\begin{lem}\label{lem:simple-trans}
  There is an $\MSO$-transduction $\Theta$ from $\sigmacol$ to $\sigmaord$
  with parameters $\PPP, \QQQ, L$ such that if $\AA$ is a
  $\sigmacol$-structure containing a simple
  pseudo-wall of order $k$ encoding a word $w$ then $\Theta(\AA)$
  contains a coloured ordered clique of order $k$ encoding $w$ and all
  $\BB\in\Theta(\AA)$ are coloured cliques encoding $w$.
\end{lem}
\begin{Proof}
  We define a transduction $\Theta := (\phivalid, \phiuniv, \phi_\sim, \phi_V,
  \phi_E, \phi_\in, \phi_{C_0}, \phi_{C_1}, \phi_{\leq})$ as
  follows. Recall that a simple pseudo-wall consists of a long path
  $L$ containing $k$ ``blue'' edges which partition the path into $k$
  sub-paths $P_1, \dots, P_k$ and a set $\QQQ$ of pairwise vertex
  disjoint paths such that for every pair $1\leq i < j \leq k$ there
  is a path in $\QQQ$ linking $P_i$ and $P_j$. 

  The parameters $\PPP, \QQQ$ will be enforced to be
  interpreted by sets of pairwise vertex disjoint paths and $L$ will
  be enforced to be a simple path. The 
  intended interpretation is that $L$ is the long path, $\PPP$ are the
  segments of $L$ without the blue edges and $\QQQ$ are the paths
  connecting the segments in $\PPP$. All this will be defined in
  $\phivalid$. But first we define the other formulas, where as usual
  we use the notation $P\in \PPP$ as shortcut for $P\subseteq \PPP
  \wedge \maxpath(P, \PPP)$.
  
  Note, that the paths in $\QQQ$ may intersect various segments $P\in
  \PPP$. Hence, in principle every vertex of a segment $P\in \PPP$ can
  also be contained in some $Q\in\QQQ$. This means that we cannot take
  the vertices of $P\in\PPP$ to represent $P$ in the transduction, as
  this would make it difficult to guarantee that $\phi_\sim$ defines
  an equivalence relation. However, as every segment $P\in\PPP$ has a non-empty
  intersection with more than one path in $\QQQ$ and the paths in
  $\QQQ$ are pairwise disjoint, every $P\in\PPP$ must contain at least
  one edge not contained in any $Q\in\QQQ$. Similarly, every
  $Q\in\QQQ$ contains an edge not contained in any other path. We will therefore take
  these unique edges to represent $P$ and $Q$, resp. 

  We define formulas $\uniqueedge_{\PPP}(e, P, \PPP, \QQQ) := e \in
  P \wedge \neg\exists Q\in\QQQ\,e\in E(Q)$ which defines an edge
  $e$ to be an edge of $P$ not contained in any path in $\QQQ$. Analogously
  we define $\uniqueedge_{\QQQ}(e, Q, \PPP, \QQQ) := e \in Q \wedge
  \neg\exists P\in\PPP\,e\in P$ and set 
  \[
     \uniqueedge(e, P) := \big(P
     \in \PPP \wedge \uniqueedge_{\PPP}(e, P)\big) \vee \big(P\in\QQQ \wedge
     \uniqueedge_{\QQQ}(e, P)\big).
  \] 

  Let $\phi_V(x) := \exists P \in \PPP\ \uniqueedge_{\PPP}(x, P)$ and 
  \[
    \phi_\sim^V(x,
    y) := \exists P \in \PPP\ \uniqueedge_{\PPP}(x, P) \wedge
    \uniqueedge_{\PPP}(y, P).
  \]
  
  We define $\phi_E(e) := \exists Q\in \QQQ \wedge
  \uniqueedge_{\QQQ}(e, Q)$ and 
  \[
    \phi_\sim^E(x,
    y) := \exists Q \in \QQQ\ \uniqueedge_{\QQQ}(x, Q) \wedge
    \uniqueedge_{\QQQ}(y, Q).
  \]
  Finally, $\phi_{\sim}(x, y) := \phi_\sim^V \vee \phi_\sim^E$ and
  $\phiuniv := \phi_V \vee \phi_E$. Note that $\phi_V$ and $\phi_E$
  define disjoint sets and therefore $\phi_\sim$ defines an
  equivalence relation on $\phiuniv$. 

  To define the colours, we set 
  \[
    \phi_{C_i}(x) := \exists P \in
    \PPP \wedge \uniqueedge_{\PPP}(x, P) \wedge \exists u \in V(P)
    \wedge u\in C_i,
  \]
  for $i\in \{0,1\}.$

  The ordering is defined by 
  \begin{eqnarray*}
    \phi_\leq(x, y)& :=& \exists p \in V(L) \exists e\in L\Big(
    \phiendpoint(p, L) \wedge p\in e \wedge  e \in B\
    \wedge\\&& 
    \forall P\subseteq L \big(\pAth(P) \wedge p\in V(P) \wedge y \in
    E(P) \rightarrow x \in E(P)\big)\Big)
  \end{eqnarray*}

  The formula $\phi_\leq(x, y)$ first defines the endpoint $p$ of the
  long path $L$ which is incident to a blue edge in $L$ (there is
  only one blue edge incident to an endpoint) and then defines $x$ to be smaller than $y$ if every
  sub-path $P$ of $L$ containing $p$ and $y$ also contains
  $x$. This defines the natural ordering on $L$ where the blue edge
  marks the left, i.e.~smaller, end.
  
  Finally, we have to define the main formula $\phivalid$ which will
  need to say that the parameters $\PPP, \QQQ, L$ indeed define a
  simple pseudo-wall as required. For this, we need to enforce the
  following requirements $\phi_1$ to $\phi_5$.
  \begin{eqnarray*}
     \phi_1 &:=& \pAth(L)\ \wedge L = B \cup R\ \wedge\\&&
     \neg \exists u\in V(L) \exists e, e' \in L \big(e\not=e'\wedge
     e\in B \wedge e'\in B \wedge u \in e \wedge u\in e'\big)\ \wedge\\&&
     \exists^{=1} p \big(\phiendpoint(p, L) \wedge
     \exists e \in L \wedge p \in e \wedge e \in B\big)
   \end{eqnarray*}
  The formula $\phi_1$ says that $L$ is a path which consists exactly
  of the red and blue edges in the structure. Furthermore, 
  in $L$ no two blue edges $e, e'\in B$ are adjacent, i.e.~between any
  two blue edges there is a red edge, and 
  $L$ has
  exactly one endpoint which is incident to a blue edge, i.e.~the
  first edge on one end is blue but the last edge is red.

  The formula 
  \[
    \phi_2 := \sodp(\PPP) \wedge \forall P \subseteq E \Big( \big(P \subseteq
    L\setminus B \wedge \maxpath(P, L\setminus B)\big) \leftrightarrow
    P\in \PPP\Big)
  \]
  says that $\PPP$ contains exactly the connected components of
  $L\setminus B$, i.e.~the segments of $L$ defined by removing the
  blue edges.

  The formula 
  \begin{eqnarray*}
     \phi_3 &:=& \sodp(\QQQ) \wedge 
     \forall P\not= P' \in \PPP \exists Q \in \QQQ\exists
     u, u'\in V(Q)\\&&
     \quad\big( \phiendpoint(u, Q) \wedge \phiendpoint(u', Q)
     \wedge u \in V(P) \wedge u'\in V(P')\big)
   \end{eqnarray*}
  says that $\QQQ$ is a set of pairwise vertex disjoint paths and that
  for any distinct pair $P, P'\in \PPP$ there is a path in $\QQQ$ linking $P$
  and $P'$, i.e.~having one endpoint in $P$ and the other in $P'$.

  Finally, we have to define that the colours are defined properly,
  i.e.~that either all vertices of a path $\PPP$ have a colour, in 
  this case it is the same colour for all, or none has a
  colour. Furthermore, we need to say that the coloured paths occur to
  the left of $L$, i.e.~if a path $P \subseteq L$ contains a coloured
  vertex then so do all $P'\subseteq L$ which are closer to the end of
  $L$ marked by a blue edge. This is formalised by the following
  formula
  \begin{eqnarray*}
    \phi_4 &:=& \forall P\in \PPP \big( (V(P) \subseteq  C_0 \vee V(P)
    \cap C_0 = \emptyset) \wedge (V(P) \subseteq  C_1 \vee V(P)
    \cap C_1 = \emptyset)\big) \ \wedge\\&&
    \exists x \in C_0 \cup C_1  \forall y \in V(L) \big( y\in C_0 \cup
    C_1 \leftrightarrow \phi_{\leq}(x, y)\big).
  \end{eqnarray*}

  The last bit we have to specify is that $B, R$ are colours of edges
  whereas $C_0, C_1$ are colours of vertices and that all colours are
  distinct. This is expressed by
  \[
    \phi_5 := \big(C_0 \subseteq V \wedge C_1 \subseteq V \wedge C_0 \cap C_1 =
    \emptyset\big) \wedge \big(R \subseteq E \wedge B \subseteq E
    \wedge R \cap B = \emptyset\big).
  \]

  Putting everything together we get
  \[
     \phivalid := \bigwedge_{i=1}^5 \phi_i.
  \]
  Now, $\phivalid$ forces the parameters $\PPP, \QQQ, L$ together to
  define a simple pseudo-wall in the structure and in this case, the
  various formulas define a coloured ordered clique encoding the same
  word as the pseudo-wall. Furthermore, the number of vertices in this
  clique is the same as the number of segments of $L$. Hence, there is
  a choice of parameters in $\AA$ where this number is the order $k$
  of the pseudo-wall. This concludes the proof. 
\end{Proof}

As before, we get the following corollary.

\begin{cor}\label{cor:simple-trans}
  Let $M$ be a non-deterministic $n^d$-time bounded Turing-machine.
  There is a formula $\phi_M\in\MSO$ such that for all words
  $w\in\Sigma^\star$, if $\AA$ is a $\sigmacol$-structure containind a
  simple pseudo-wall of order $|w|^d$
  encoding $w$, then $\AA\models \phi_M$ if, and only if, $M$
  accepts $w$.  Furthermore, the formula $\phi_M$ can be constructed
  effectively from $M$.  

  The same holds if $M$ is an alternating
  Turing-machine with a bounded number of alternations, as they are
  used to define the polynomial-time hierarchy.
\end{cor}

\subsection{Complex Pseudo-Walls}

We will now define a transduction from complex pseudo-walls to ordered
coloured cliques. As complex pseudo-walls are more complex than simple
ones, we will do so in two steps. We first exhibit a transduction with
parameters $\PPP, \QQQ, L$ that will enforce $\PPP, \QQQ, L$ to
satisfy the requirements of a complex pseudo-wall and will then
generate the intersection graph of $\PPP$ and $\QQQ$ where vertices
are suitably coloured as prescribed by the definition of a complex
pseudo-wall. By definition of a complex pseudo-wall, this intersection
graph contains a topological clique-minor. The second transduction,
therefore, will generate a coloured ordered clique from this clique
minor.

For the first step, we define a transduction 
\[
  \Theta_1 := (\phivalid, \phiuniv, \phi_\sim, \phi_V, \phi_E,
  \phi_\in, \phi_{C_0}, \phi_{C_1}, \phi_\leq) 
\]
from $\sigmacol$ to $\sigmaord$ with parameters $\PPP,
\QQQ, L$ as follows. 

Again, $\phivalid$ will ensure that $\PPP, \QQQ$ are interpreted by
sets of pairwise vertex disjoint paths, so we will use previous
notation such as $Q\in\QQQ$. 

Recall that in the intersection graph $\III(\PPP, \QQQ)$ the vertices
are the paths in $\PPP$ and $\QQQ$ and an edge exists between $P\in
\PPP$ and $Q\in \QQQ$ if they intersect. Hence, in $\Theta_1$ we will
represent a path $P\in\PPP$ by its unique edges (see the previous
subsection) and an edge $\{P, Q\}$ by the vertices in the intersection
of $P$ and $Q$.

Thus, we define $\phi_V(x) := x \in E \wedge \big(\exists P\in \PPP
\uniqueedge_{\PPP}(x, P) \vee \exists Q\in \QQQ \uniqueedge_{\QQQ}(x, Q)\big)$ and
\[
  \phi_\sim^V(x, y) :=
  \begin{array}{l}
    \exists P\in\PPP (\uniqueedge_{\PPP}(x, P) \wedge \uniqueedge_{\PPP}(y,
    P))\ \vee\\
    \exists Q\in\QQQ ( \uniqueedge_{\QQQ}(x, Q) \wedge \uniqueedge_{\QQQ}(y,
    Q)).
\end{array}
\]
Furthermore, we define $\phi_E(x) := \exists
P\in\QQQ \exists Q\in\QQQ (x\in V(P)\cap V(Q))$ and 
\[
  \phi_\sim^E(x, y)
  := \exists P\in\PPP \exists Q\in\QQQ x\in V(P) \cap V(Q) \wedge y\in
  V(P) \cap V(Q).
\]
 Finally, we define $\phiuniv(x) := \phi_V(x) \vee
\phi_E(x)$ and $\phi_\sim(x, y) := \phi_\sim^V(x, y) \vee
\phi_\sim^E(x, y)$. 

It is easily seen that $\phi_\sim^V$ and $\phi_\sim^E$ define
equivalence relations on the sets defined by $\phi_V(x)$ and
$\phi_E(x)$, resp., and as these sets are disjoint also on the set
defined by $\phiuniv$.

Let $\phi_\in(x, e) := $
\[
  \exists P \in \PPP \exists Q\in  \QQQ \Big( e\in V(P)
  \wedge e\in V(Q) \wedge \big(\uniqueedge_{\PPP}(x, P) \vee \uniqueedge_{\QQQ}(x,
  Q)\big)\Big)
\]
The formula states that $e$ is a vertex in the intersection of a path
$P\in\PPP$ and a path $Q\in \QQQ$, and therefore representing an edge
between $P$ and $Q$,
and $x$ is a unique edge of one of the two paths and hence represents
a vertex for $P$ or $Q$. 

We define the colours $C_0$ and $C_1$ next. Here, we give a vertex $P\in\PPP$
the colour $C_i$ if the (uniquely defined) endpoint of the path $P$ in
the long path
$L$ is in $C_i$. Recall that in a complex pseudo-wall, every path $P$
intersect $L$ in exactly one of its endpoints. The colours are
therefore defined by the formulas
\[
  \phi_{C_i}(x) := \exists P\in\PPP \big(\uniqueedge_{\PPP}(x, P) \wedge \exists y
  \in V(P) \cap V(L) \wedge y\in C_i\big),
\]
where $i\in \{ 0,1\}$. Note that we do not need to state that $y$ is an
endpoint of $P$ as $P$ can intersect $L$ only once. 

Finally, we define an ordering on the vertices constituted by paths in
$\PPP$. The ordering we aim at is the natural ordering given
by $L$, where a path $P$ is smaller than a path $P'$ if the endpoint
of $P$ in $L$ is closer to the blue edge in $L$ than the endpoint of
$P'$ in $L$.

\begin{eqnarray*}
  \phi_\leq(x, y) &:=& \phi_V(x) \wedge \phi_V(y) \wedge \exists P, P'
  \in \PPP  \uniqueedge_{\PPP}(x, P) \wedge \uniqueedge_{\PPP}(y, P') \wedge\\&&
  \exists u,u'\big( u\in V(P)\cap V(L) \wedge u' \in V(P') \cap V(L)
  \wedge \phi_{\leq_L}(u,u') \big),
\end{eqnarray*}
where 
\begin{eqnarray*}
  \phi_{\leq_L}(u,u') &:=& \exists e\in L \Big(e\in B \wedge \exists
  y\in V(L) \big(\phiendpoint(y, L) \wedge y\in e\wedge\\&&
  \forall P\subseteq L\big(
  \pAth(P)\wedge y\in V(P) \wedge u'\in V(P) \rightarrow u\in V(P)\big)\big)\Big).
\end{eqnarray*}

The last part of $\Theta_1$ to be defined is $\phivalid$. 
Again we will do this in various steps.
\begin{eqnarray*}
  \phi_1 &:=& \pAth(L)\ \wedge (L = B \cup R) \wedge \exists^{=1}
  e \in B \wedge \exists^{=1}
  e \in B \big( \exists u (\phiendpoint(u, L) \wedge u\in e)\big)
\end{eqnarray*}

The formula says that $L$ is a path comprising all red and blue edges
and that there is exactly one blue edge and this is the  first on the path.

The next formula $\phi_2$ says that only vertices on $L$ are coloured
and that no vertex has two colours.
\[
   \phi_2 := C_0 \cup C_1 \subseteq V(L) \wedge C_0 \cap C_1 = \emptyset
\]
Finally, we need to say that $\PPP$ and $\QQQ$ are sets of pairwise
disjoint paths and that each path in $\PPP$ has exactly one endpoint
on $L$ and is otherwise vertex disjoint from $L$. This is expressed in
the next formula.
\[
   \phi_3 := \sodp(\PPP) \wedge \sodp(\QQQ) \wedge \forall
   P\in \PPP\big(\exists x \phiendpoint(x, P) \wedge V(P)\cap V(L) = \{ x\}\big)
\]

Now, we set $\phivalid := \phi_1 \wedge \phi_2 \wedge \phi_3$.

Let $\AA$ be a complex pseudo-wall and let $\PPP, \QQQ, L$ be sets
of edges such that $(\AA, \PPP, \QQQ, L) \models \phivalid$. Hence,
$\PPP$ and $\QQQ$ are sets of vertex dispoint paths. 

Let
\begin{eqnarray*}
  U &:=& \{ [x]_{/\phi_\sim(\AA)} \st x\in \phiuniv(\AA) \},\\ 
  V &:=& \{ [x]_{/\phi_\sim(\AA)} \st x\in \phi_V(\AA) \},\\
  E &:=& \{ [x]_{/\phi_\sim(\AA)} \st x\in \phi_E(\AA) \} \text{ and }  \\
  \in_U &:=& \{ ([x]_{/\phi_\sim(\AA)}, [y]_{/\phi_\sim(\AA)}) \st (x, y)\in \phi_\in(\AA)\}. 
\end{eqnarray*}
By construction,
$( U, V, E, \in_U)$ is isomorphic to the intersection graph
$\III(\PPP, \QQQ)$ of $\PPP$ and $\QQQ$. Furthermore, $C_i := \{ \{
[x]_{/\phi_\sim(\AA)} \st x\in \phi_{C_i}(\AA) \}$, for $i\in \{
0,1\}$, define colours of vertices in $V$ and $\leq_V := \{  [(x,
y)]_{/\phi_\sim(\AA)} \st (x, y)\in \phi_\leq(\AA)\}$ defines a linear
order on the subset of the vertices 
of $\III(\PPP, \QQQ)$ corresponding to paths in $\PPP$.

By definition, if $\AA$ is a complex pseudo-wall, then we can choose
$\PPP$ and $\QQQ$ so that $\III(\PPP, \QQQ)$ contains a topological
clique-minor such that every branch set contains at most one coloured
vertex and all coloured vertices occur in a branch set. This is
clearly not the case for all choices of $\PPP, \QQQ, L$ satisfying
$\phivalid$, but for our purposes it will be enough to know that there
is one such choice. 

We will now exhibit a second transduction 
\[
  \Theta_2 := (\phivalid,
  \phiuniv, \phi_\sim, \phi_V, \phi_E, \phi_\in, \phi_{C_0}, \phi_{C_1},
  \phi_\leq)
\]
with parameters $X, F, T$ which
defines a coloured ordered clique encoding the same word as $\AA$ in
some structures in $\Theta_1(\AA)$. Here we benefit from the fact that
we only need to define topological minors, which makes the next
transduction easy to define. The parameters $X, F, T$ have the
following intuitive meaning. By definition of topological minors,
if $K_n$ is a topological minor of a graph $G \in \Theta_1(\AA)$ then
there are $n$ vertices $u_1, \dots, u_n$ in $G$ and for all $1\leq
i<j\leq n$ a path $P_{i,j}$ between $u_i$ and $u_j$ such that if
$\{a,b\}\not=\{i,j\}$ then $P_{a,b}$ and $P_{i,j}$ are internally
vertex disjoint (they have an endpoint in common if
$\{a,b\}\cap\{i,j\} \not=\emptyset$). The parameter $X$ will denote
the set $\{ u_1, \dots, u_n\}$ and $F$ will be the union
$\bigcup_{i < j} E(P_{i,j})$. Hence, the graph defined by $\Theta_2$
will have $X$ as vertex set and the individual $P_{i,j}$ as edges. To
define the colours of the vertices in $X$ we need the last parameter
$T$. $T$ will contain exactly
one edge of each path $P_{i,j}$. This will act as a separator: with
every $x\in X$ we associate the set of all vertices on the paths
$P_{i,j}$ emerging from $x$ up to the edge in $T$. We will then say
that for each $x$ this set contains exactly on coloured vertex and we
will take the colour of this vertex as colour of $x$. 

$\Theta_2$ is now formally defined as follows. To define the vertices
let $\phi_V(x) := x \in X$ and $\phi_\sim^V(x, y) := x=y$. 
To define
edges we first need some preparation. 

Let 
\[
\textit{F-path}(P, x, x') :=
\begin{array}{l}
P\subseteq F\wedge \pAth(P) \wedge
\phiendpoint(x, P) \wedge \phiendpoint(x', P)\ \wedge\\\neg \exists y\in
X(y\in V(P) \wedge y\not=x \wedge y\not= x').
\end{array}
\]
 The formula says that
$P$ is a path whose edges are all from $F$, whose end points are
$x$ and $x'$ and which contains no other vertex from $X$. 
Let 
\[
  \textit{mp}(P, F, X) := P\subseteq F \wedge \exists x, x'\in X
  \textit{F-path}(P, x, x').
\] 
The formula says that $P$
is a path with edge set in $F$ connecting two vertices $x, x'\in X$. 

Let $\phi_E(e) := \exists P\subseteq F\big( \textit{mp}(P, F, X) \wedge
e\in P\big)$ and 
$\phi_\sim^E(x, y) := \exists P\subseteq F \big(\textit{mp}(P,
F, X) \wedge x\in P \wedge y\in P\big)$. As mentioned above, we will
represent edges by paths $P_{i,j}$ in $F$ between vertices in
$X$. $\phi_E(e)$ says that $e$ is an edge of such a path and
$\phi_\sim^E(e, e')$ defines $e$ and $e'$ to be equivalent if they
occur on the same path in $F$. 
As usual, $\phiuniv(x) := \phi_V(x) \vee \phi_E(x)$. 

We now define the colours $C_0, C_1$. First, let
\[
  \textit{branch-set}(x, y) := x\in X \wedge \exists Q\subseteq F \big(
  \maxpath(Q, F\setminus T) \wedge x,y\in V(Q)\big).
\]
The formula defines for given $x\in X$ the set of all vertices that
can be reached from $x$ by a path with edges of $F$ not containing
any edge from $T$. We can now define $\phi_{C_i}(x) := \exists
y\big(\textit{branch-set}(x, y) \wedge y\in C_i\big)$, for $i\in\{0,1\}$.

Finally, we define $\phi_\leq(x, y) := x\leq y$.

The last part of $\Theta_2$ left to be defined is $\phivalid$. Here we
must say that $X$ and $F$ indeed induce a topological clique-minor as
indicated above and that $T$ is a separator containing one edge from
each path linking two vertices from $X$. 

We first use the formula 
\[
  \phi_0 := X\subseteq V \wedge F\subseteq
E\wedge T\subseteq F \wedge C_0\cap C_1 = \emptyset
\]
 to say that the parameters are of the right
type.

The formula $$\phi_1 := \forall x, x'\in X \Big( x\not= x'
\rightarrow \exists P \textit{F-path}(P, x, x') \wedge \forall Q
\big(\textit{F-path}(Q, x, x') \rightarrow P=Q\big)\Big)$$ says that any two
distinct vertices in $X$ can be connected by a path in $F$ and that this is
unique.

The formula 
\[
\phi_2 := \forall e (e\in F \rightarrow \exists x, x'\in
X\exists P\subseteq F( \textit{F-path}(P, x, x') \wedge e\in P))
\]
 says
that every edge of $F$ occurs on a path in $F$ between two vertices of
$X$. 

The formula 
\[
  \phi_3 := \forall x, y, x', y' \in X\ x\not= x'
  \rightarrow \left(
  \begin{array}{c}
  \exists P, P'\subseteq F \big(\textit{F-path}(P, x, y)
  \wedge \textit{F-path}(P', x', y')\big)\ \wedge\\
  (y\not= y' \rightarrow V(P')\cap V(P) = \emptyset)\ \wedge\\
  (y= y' \rightarrow V(P')\cap V(P) = \{ y\})
  \end{array}\right)
\]
says that if $x, x', y, y'$ are distinct vertices in $X$ then the
paths $P, P'$ linking $x$ to $y$ and $x'$ to $y'$, resp., are pairwise
vertex disjoint and if $x, x', y, y'$ are such that $y=y'$ but
$x\not=x'$ then the two paths only have $y$ in common. 

The formulas $\phi_0, \phi_1, \phi_2, \phi_3$ together imply that $(X, F)$
induce a topological clique minor as required. 
We next define a formula saying that $T$ is as required, i.e.~$T$
contains one edge from each path connecting two vertices in $X$ and
that every edge of $T$ is contained in such a path.
\[
  \phi_4 :=
  \begin{array}{l}
    \forall x, x'\in X\exists P \subseteq F
    \big(\textit{F-path}(P, x, x') \wedge \exists^{=1} e\in P (e \in
    T)\big)\ \wedge\\
    \forall e\in T \exists x, x'\in X \exists P\subseteq
    F\ (\textit{F-path}(P, x, x') \wedge e\in P)

\end{array}
\]

What is left to define
are the colours and that all vertices in $X$ can be linearly ordered
by $\leq$. The latter is easily defined by $\phi_5 := \forall x(x\in
X\rightarrow x\leq x)$. 

The formula
\[
  \phi_6 := \forall x\exists^{\leq1} y \big( \textit{branch-set}(x, y)
  \wedge y\in C_0\cup C_1) \wedge \forall c\in C_0 \cup C_1 \exists
  x\in X
  \textit{branch-set}(x, c)
\]
says that every branch set contains at most one coloured vertex and
every coloured vertex is contained in a branch set. 

Finally, we need to say that the vertices $x$ whose branch sets
contain a coloured vertex are the smallest with respect to
$\leq$. This is stated by the formula
\[
  \phi_7 := \exists x\in X\big(\phi_{C_0}(x) \vee \phi_{C_1}(x) \wedge
  \forall y\in X (\phi_{C_0}(y) \vee \phi_{C_1}(y) \rightarrow y\leq x\big).
\]

Let $\phivalid := \bigwedge_{i=0}^7 \phi_i$. 

Now, if $\AA$ is a $\sigmaord$-structure and $X\subseteq V^\AA$ and
$F, T\subseteq E^\AA$ then $(\AA, X, F, T)\models \phivalid$ if, and
only if, $(X, F)$ determines a topological clique-minor where the
branchsets are the components of $F\setminus T$. Furthermore,
$\Theta(\AA, X, F, T)$ is a coloured ordered clique encoding the same
word as $\AA$ and all $\BB\in\Theta_2(\AA, X, F, T)$ are coloured
cliques encoding the same word as $\AA$. 

The interpretations $\Theta_1$ and $\Theta_2$ together yield the
desired transformation of complex pseudo-walls to coloured ordered
grids as stated in the following lemma. 

\begin{lem}\label{lem:complex-trans}
  If $\AA$ is a $\sigmacol$-structure containing a complex pseudo-wall of order $k$ encoding a word $w$
  then $\Theta_2(\Theta_1(\AA))$ contains a coloured ordered clique of
  order $k$ encoding $w$.  Furthermore, every
  $\BB\in\Theta_2(\Theta_1(\AA))$ is a coloured ordered clique
  encoding $w$. 
\end{lem}

Again, using Corollary~\ref{cor:TM-clique} and the formula translation
provided by the transductions, we get the following corollary. 

\begin{cor}\label{cor:complex-trans}
  Let $M$ be a non-deterministic $n^d$-time bounded Turing-machine.
  There is a formula $\phi_M\in\MSO$ such that for all words
  $w\in\Sigma^\star$, if $\AA$ is a $\sigmacol$-structure containing  a complex pseudo-wall of order $|w|^d$
  encoding $w$, then $\AA\models \phi_M$ if, and only if, $M$
  accepts $w$.  Furthermore, the formula $\phi_M$ can be constructed
  effectively from $M$.  

  The same holds if $M$ is an alternating
  Turing-machine with a bounded number of alternations, as they are
  used to define the polynomial-time hierarchy.
\end{cor}

The following result combines everything we need from this section
later on. 

\begin{cor}\label{cor:pw-trans}
  Let $M$ be a non-deterministic $n^d$-time bounded Turing-machine.
  There is a formula $\phi_M\in\MSO$ such that for all words
  $w\in\Sigma^\star$, if $\AA$ is a $\sigmacol$-structure containing
  either a simple or complex pseudo-wall of order $|w|^d$
  encoding $w$, then $\AA\models \phi_M$ if, and only if, $M$
  accepts $w$.  Furthermore, the formula $\phi_M$ can be constructed
  effectively from $M$.  

  The same holds if $M$ is an alternating
  Turing-machine with a bounded number of alternations, as they are
  used to define the polynomial-time hierarchy.
\end{cor}
\begin{Proof}
  Note that in a complex pseudo-wall there is at most one blue edge
  $e\in B$ whereas a simple pseudo-wall always contains more than
  one. So we can easily distinguish in first-order logic between
  simple and complex pseudo-walls.

  Now let
  \[
  \phi_M := \big(\exists^{=1} e\in B \wedge \phi^c_M \big)\vee
  \big(\exists^{\geq 2} e\in B \wedge \phi^s_M\big),
  \]
  where $\phi^c_M, \phi^s_M$ are the formulas from Corollary~\ref{cor:complex-trans}
  and~\ref{cor:simple-trans} respectively. 

  Then Corollary~\ref{cor:simple-trans}
  and~\ref{cor:complex-trans} imply that $\phi_M$ is indeed true in
  $\AA$ if, and only if, $M$ accepts $w$.
\end{Proof}

\section{Putting it all together}
\label{sec:main}

In this section we conclude the proof of Theorem~\ref{theo:main} by
combining the results obtained in Section~\ref{sec:pseudo-walls}
and~\ref{sec:msodef}. 
More precisely, we will first show the following lemma, which implies
Part $2$ of the theorem.

\begin{lem}
  Let $\CCC$ be a class of
  $\sigmacol$-structures closed under colourings. 

   If the tree-width of $\CCC$ is $(\log^c, p)$-unbounded, for
    some  $c>d\cdot 84$ and polynomial $p$ of degree $d$, 
    then $\MC(\MSO_2, \CCC)$ is not in XP and hence not fixed-parameter tractable unless
    \textsc{Sat} can be solved in sub-exponential time. 
\end{lem}
\begin{Proof}
  We show that if $\pMC(\MSO, \CCC)$ is in XP then the propositional
  satisfiability problem SAT, i.e.~the problem to decide for a formula
  of propositional logic if it has a satisfying assignment, can be
  solved in sub-exponential time.

  Let $w$ be a propositional logic formula. We can decide whether
  $w$ is satisfiable as follows.

  We first construct a $\sigmacol$-structure $\AA\in\CCC$ of
  tree-width between $c'\cdot m^{84}$ and $c\cdot m^{d\cdot 84}$,
  where $c$ is the constant from Theorem~\ref{theo:pseudo-wall} and
  $c' := \sqrt[d]{c}$.  Furthermore, $\tw(\AA) > \log^{d\cdot
    84+\delta} |\AA|$, for some $\delta>0$. Let $G$ be the
  $\sigmai$-reduct of $\AA$, i.e.~the underlying uncoloured graph of
  $\AA$.

  It follows that
  \[
  \begin{array}{crcl}
    & c\cdot m^{d \cdot 84} &>& \log^{d\cdot 84+\delta} |G|\\
    \Iff & c'' \cdot m^{\frac1y} & > & \log |G|\\
    \Iff & |G| &<& 2^{c''\cdot m^{\frac1y}}
  \end{array}
  \]
  for some $c''$ and $y> 1$.

  By Theorem~\ref{theo:pseudo-wall}, as
  $\big(\frac{\tw(G)}{\sqrt{\log{\tw(G)}}}\big)^{\frac13} \geq c\cdot
  m^{14}$, we can compute in polynomial time a $\sigmacol$-expansion
  $\BB\in\CCC$ of $G$ containing a pseudo-wall encoding $w$ with power
  $2$.

  Clearly, SAT can be decided by a non-deterministic Turing-machine
  $M$ running in time quadratic in the size of the input.  Hence, by
  Corollary~\ref{cor:pw-trans}, there is a formula $\phi_M$, depending
  only on $M$, such that $\BB\models\phi_M$ if, and only if, $M$
  accepts $w$ if, and only if, $w$ is satisfiable.

  By Definition~\ref{def:strongly}, we can construct $\AA$, and hence
  $G$, in time at most $2^{(c\cdot m^{84})^\epsilon}$ for some
  $\epsilon<1$. By Theorem~\ref{theo:pseudo-wall}, $\BB$ can be
  constructed in time polynomial in the size of $G$ and thus in time
  $2^{d'\cdot (c\cdot m^{84})^\epsilon}$, for some constant $d'$.

  Suppose now that $\pMC(\MSO, \CCC)$ is in XP, i.e.~given $G\in\CCC$
  and $\phi\in\MSO$, we can decide $G\models \phi$ in time
  $|G|^{f(|\phi|)}$, for some computable function $f\st\N\rightarrow
  \N$. Hence, we can decide $\BB\models\phi_M$ in time
  $|\BB|^{f(|\phi_M|)}$ and thus in time $|G|^{f(|\phi_M|)}$.  But
  $|G|^{f(|\phi_M|)} < 2^{f(|\phi_M|)\cdot c''\cdot m^{\frac1y}} \in
  2^{o(|w|)}$. Hence, we can decide whether $w$ is satisfiable in
  sub-exponential time. This concludes the proof of the lemma.
\end{Proof}

The lemma clearly implies Part $2$ of Theorem~\ref{theo:main}. 
Unsing any other language in the polynomial-time hierarchy instead of SAT we get the first part
by exactly the same argument. 
This concludes the proof of Theorem~\ref{theo:main}.

\section{Conclusion and Further Work}
\label{sec:conclusion}
In the previous section we have seen that if $\CCC$ is closed under
colourings and its tree-width is not bounded logarithmically, then
$\MC(\MSO, \CCC)$ is not in XP unless SAT can be solved in
sub-exponential time.
What this shows is that Courcelle's theorem cannot be extended beyond logarithmic tree-width in
its full generality.

The proof given in this paper shows that in order to apply our theorem
to a class $\CCC$, its tree-width must be $(\log^c , p)$-unbounded for
$c>84+d$, where $d$ is the degree of $p$. Using slightly more complex
algorithmic results from \cite{KreutzerTaz10} this bound can be
improved slightly to $c>48+d$. Furthermore, it is possible to reduce
the numbers of colours needed to two binary and one unary relation symbols.

Our result refers to $\MSO_2$, i.e.~monadic second-order logic with
quantification over sets of edges. If we restrict ourselves to
$\MSO_1$ then this logic becomes tractable on the much larger class of
graphs of small \emph{clique-width}.

\begin{thm}[\cite{CourcelleMakRot00}]
  Let $\CCC$ be a class of graphs of bounded clique-width. Then $\MC(\MSO_1,
  \CCC)$ is fixed-parameter tractable.
\end{thm}

It would be interesting to study classes of graphs closed under
taking induced sub-graphs which have unbounded clique-width. We therefore put
forward the following conjecture.

\begin{conj}
 If $\CCC$ is a class of graphs whose clique-width is 
poly-logarithmically  unbounded and which is closed under induced sub-graphs, then
$\MC(\MSO_1, \CCC)$ is not fixed-parameter tractable.
\end{conj}

However, so far no analogue of grid-like minors for clique-width exists and
therefore more research on obstructions for clique-width is needed to prove
this conjecture.

\bibliographystyle{plain}

\end{document}